\documentclass[twoside]{elsart} 
\journal{Astronomy \& Astrophysics}
\usepackage[dvips]{graphicx}
\usepackage[dvips]{epsfig}
\usepackage[dvips]{color}
\usepackage{natbib}
\usepackage{amssymb}  
\newcommand{\ee}{\end{equation}}

\begin{document}

\def\deg{^{\circ}}
\def\be{\begin{equation}}
\def\ee{\end{equation}}
\def\dl{\Phi^{\mbox{DL}}}
\def\stl{\Phi^{\mbox{SL}}}
\def\sdl{\Phi^{\mbox{SDL}}}
\def\ds{\Phi^{\mbox{DS}}}
\def\ps{\Phi^{\mbox{PS}}}
\def\diffflux{\mbox{GeV}^{-1}\,\mbox{cm}^{-2}\,\mbox{s}^{-1}\,\mbox{sr}^{-1}}
\def\pointflux{\mbox{GeV}^{-1}\,\mbox{cm}^{-2}\,\mbox{s}^{-1}}
\def\diffunits{\mbox{GeV cm}^{-2}\,\mbox{s}^{-1}\,\mbox{sr}^{-1}}
\def\pointunits{\mbox{GeV cm}^{-2}\,\mbox{s}^{-1}}
\def\en{E_{\nu}}
\def\eg{E_{\gamma}}
\def\ep{E_{p}}
\def\epb{\epsilon_{p}^{b}}
\def\enb{\epsilon_{\nu}^{b}}
\def\enbG{\epsilon_{\nu,GeV}^{b}}
\def\enbM{\epsilon_{\nu,MeV}^{b}}
\def\ens{\epsilon_{\nu}^{s}}
\def\ensG{\epsilon_{\nu,GeV}^{s}}
\def\egb{\epsilon_{\gamma}^{b}}
\def\egbM{\epsilon_{\gamma,MeV}^{b}}
\def\g25{\Gamma_{2.5}}
\def\lumi{L_{\gamma}^{52}}
\begin{frontmatter}
\title{On the origin of ultra high energy cosmic rays:\\
 {\large Subluminal and superluminal relativistic shocks}}~\vspace{-1.5cm}
\author[ath,dort,cor]{Athina Meli}
\author[dort,swed,cor]{, Julia K.~Becker}
\author[london]{, John J.~Quenby }
\corauth[cor]{{\scriptsize Corresponding authors:\\ameli@phys.uoa.gr,
    phone: +30-210-727-6908\\julia.becker@physics.gu.se, phone: +46-31-7723190}}
\address[ath]{Department of Physics, National University of Athens, Panepistimiopolis
Zografos 15783, Greece}
\address[dort]{Institut f\"ur Physik, Universit\"at Dortmund, 44221 Dortmund, Germany}
\address[swed]{Institutionen f\"or Fysik, G\"oteborgs Universitet, 41296 G\"oteborg, Sweden}
\address[london]{High Energy Physics Group, Blackett Laboratory, Imperial College London, Prince Consort Road SW7 2BZ, UK}

\date{\today}~\vspace{-1.5cm}

\begin{abstract}
The flux of ultra high energy cosmic rays (UHECRs) at $E>10^{18.5}$~eV is believed to
arise in plasma shock environments in extragalactic sources. In this paper, we
present a systematic study of particle acceleration by relativistic shocks, in particular concerning
the dependence on bulk Lorentz factor and the angle between the magnetic field and the shock
flow. For the first time, simulation results of super- and subluminal shocks with boost factors up to
$\Gamma=1000$ are investigated and compared systematically. 
While superluminal shocks are shown to be inefficient at the highest energies ($E>10^{18.5}$~eV), 
subluminal shocks may provide particles up to $10^{21}$~eV, limited
only by the Hillas-criterion.
For the subluminal case, we find that mildly-relativistic shocks, thought to occur in jets
of Active Galactic Nuclei (AGN, $\Gamma\sim 10-30$) yield energy spectra of
$dN/dE\sim E^{-2}$. Highly-relativistic shocks expected in Gamma Ray
Bursts (GRBs, $100<\Gamma<1000$), on the other hand, have spectra as flat as $E^{-1.5}$.
The model results are compared to the measured flux of cosmic rays (CRs) at the
highest energies and  it is shown that, while AGN spectra are well-suited, GRB spectra are too flat to explain the
observed flux. The first evidence of a correlation between the cosmic ray flux
above $5.7\cdot 10^{10}$~GeV and the distribution of AGN by Auger are
explained by the model.
Neutrino production is expected in GRBs, either in mildly 
or highly relativistic shocks and although these sources are excluded as 
the principle origin of UHECRs, superluminal shocks in particular may be 
observable via neutrino and photon fluxes, rather than as protons.
\end{abstract}

\end{frontmatter}


\section{Introduction \label{introduction}}

The observation of the energy spectrum of ultra high energy cosmic rays (UHECRs)
indicates the presence of an extragalactic component at
$E>10^{18.5}$~eV. Active Galactic Nuclei (AGN) and Gamma Ray Bursts (GRBs) seem
to be the two most promising source candidates for the production of charged cosmic rays
(CRs). Work in the late 1970s by a number 
of authors, e.g. \cite{krymskii77,bell78a,bell78b}, who based their 
idea on the original Fermi acceleration mechanism, first presented by~\cite{fermi49, fermi54},
established the basic mechanism of particle diffusive acceleration in 
non-relativistic shocks. In this mechanism, individual particles 
are accelerated in a collisionless magnetised plasma by scattering off magnetic
irregularities to recross a shock front many times.  
Since then, considerable analytical and 
numerical investigations have been performed, but more questions need to be answered concerning the 
acceleration mechanism at relativistic shock speeds.

In this work, we will present a series of simulation studies varying
the shock velocity and the magnetic field inclination to the shock normal, 
applying various particle scattering media, fixed in the plasma flow frame, 
 aiming to provide a more refined determination of 
the possible spectra which could result. All shocks under investigation are taken to be oblique, 
so that the magnetic field is neither aligned with nor strictly perpendicular to the shock front
normal. Parallel shock simulation studies were given in \cite{meli_quenby03_1}.
Monte Carlo calculations will be performed in the relativistic shock environments 
believed to occur in AGN and GRB jets. The resulting very high energy CR spectra and the subsequent
multiwavelength radiation observation from candidate sources, such as AGN and GRBs,
will be mentioned.

Previous work by  \cite{meli_quenby03_1,meli_quenby03_2}, on simulation studies 
for subluminal and superluminal shocks did not establish the relationship between spectral slope,
shock inclination angles, shock velocity and particle scattering 
model which will be considered here.
These past studies amongst others investigated the 'traditional'
large angle scattering model and the pitch angle diffusion model of particles, but only for 
the extreme limiting case of $\delta \theta \leq 1/\Gamma$. 
In the present work we will allow for
pitch angle scattering to vary between $1/\Gamma \leq \delta \theta \leq 10/\Gamma$ to correspond to
a variety of scattering wave models.
We will establish the spectral index dependence 
on the $\Gamma$ of the shock. The shock obliquity will be varied while using a series of 
high velocity plasma flows, ranging from $\Gamma=10-1000$.
The contribution of the high energy CRs from 
AGN and GRBs to the observed diffuse CR spectrum will be discussed, based upon the simulated
spectra. 

In Section~\ref{simulations}, 
details of the simulation are discussed. Section~\ref{results} presents the resulting spectra 
emphasising the spectral dependence on the relativistic flow Lorentz 
factor. The calculated spectra are used to estimate a
possible contribution of AGN and GRBs to the observed diffuse spectrum of
charged CRs. The implications for high energy neutrino and photon
emission is discussed in Section~\ref{neutrinos:sec}. Finally, in Section~\ref{summary} the results are summarised.
\subsection{Source Candidates for relativistic shocks}

The energy budget available for UHECRs production in GRBs 
is approximately  $2\times 10^{44}$~erg/Mpc$^{3}$/yr, as concluded from the
observed  gamma-ray GRB release rate. This energy budget corresponds to  
the required energy release rate of $>10^{18.5}$ eV of UHECRs and is
based on an assumed star formation rate (SFR)
history~\citep{vietri95,waxman_emax} with about $1$~burst/Gpc$^3$/yr at $z=0$.
The typical luminosity of AGN, $L\sim 10^{42}-10^{47}$~erg/s is also
sufficient to explain the UHECR flux under the assumption that AGN follow the
SFR. For example, \cite{Moran01} find an average luminosity from 
an estimated 95 X-ray active AGN within 60 Mpc of $4.8\times 10^{44}$ erg/s, this 
distance being the cosmic ray absorption horizon at about $2\times10^{20}$ eV.
Within the regions of local space accessible to cosmic ray diffusion, the energy supply
over a Hubble time is $6.6\times10^{-16}$ erg/cm$^{3}$. By comparison the GRB
supply is $6.7\times10^{-20}$ erg/cm$^{3}$.    
 With GRBs as the most luminous transient objects in the sky and AGN as
the most luminous permanent ones, these two source classes are the best
candidates for the acceleration of UHECRs, following the arguments of~\cite{hillas}. 

The observation of electron synchrotron radiation in the radio regime indicates
that mildly-relativistic shocks of boost factors $\Gamma\approx 10-30$ are
present in the jets of AGN ~\citep{BiermannStrit87,falcke2}. The photon spectrum of 
AGN is broadband, ranging from radio up to TeV emission in the case of optically 
thin sources. Assuming that hadrons are accelerated along with the electrons in the jet, 
AGN are good candidates to be responsible for at least a significant fraction of the
extragalactic component of the CR flux. 

The observation of the highly variable, prompt GRB photon spectra at
soft photon energies ($E_{\gamma}>100$~keV) indicate the acceleration
of electrons in highly relativistic shocks of boost factors around
$100<\Gamma<1000$, see~\cite{halzenhooper02} and references
therein. However, mildly relativistic internal shocks may also occur
within the GRB plasma flow.
Protons are believed to be accelerated up to $\sim 10^{21}$~eV as
discussed by~\cite{waxman_emax}. The total release of
electromagnetic energy by GRBs has led to the suggestion that the CR
spectrum above the ankle ($E>10^{18.5}$~eV) can be explained by
GRBs~\citep{vietri95}. 
\subsection{Maximum energy \label{emax}}
A basic physical limitation to the maximum energy of the accelerated particles is the
size of the acceleration region and the magnetic field present. 
\cite{Parker58a,Parker58b,Parker58c, Parker66} first discussed this CR energy 
limitation within the solar system.
Later on \cite{hillas} extended the CR energy limit argument
to a number of astrophysical sources.
Particles must escape once the gyration radius exceeds the source radius.
The maximum energy is then given as
\be
E_{\max}^{18}=\beta_s\cdot Z\cdot B_{\mu G}\cdot L_{kpc}\,.
\ee
Here, $E_{\max}^{18}:=E_{\max}/(10^{18}$~eV) is the maximum energy that can
be achieved, $\beta_s=V_s/c$ and is equal to 1 for the oblique shock conditions \citep{jokipii87}, $Z$ is the charge 
of the accelerated particle in units of the charge of the electron, $e$. Furthermore, $B_{\mu G}:=B/(1\,\mu$G)
is the magnetic field of the acceleration region in units of $1\,\mu$G and $L_{kpc}:=L/(1$~kpc) is
the size of the acceleration region in units of $1$~kpc.
As discussed by~\cite{hillas}, AGN cores, radio galaxy lobes, and hot spots,
 are promising candidates for the acceleration of the highest energy events. A second criterion of Hillas
that the proton synchrotron loss time should not be less than the acceleration time,
is easily met by proton shock acceleration if the shock is relativistic since the acceleration time is
\be
\tau_{acc}=\frac{3\times10^{3}XE_{18}}{B_{\mu G}}~~ \rm{year} \,,
\ee
where the scattering mean free path is $X$ times the Larmor radius and the synchrotron loss time is
\be
\tau_{synch}=\frac{1.4\times10^{14}}{E_{18}B_{\mu G}^{2}}~~\rm{year} \,.
\ee

As mentioned before, boost factors in AGN are typically assumed to be around $\Gamma\sim 10$, although in
special cases, the boost factor can reach up to $\Gamma\sim 30-40$.
It is, however, possible that the shock itself happens within a general relativistic
environment  \citep{BiermannStrit87} and so the shock is not necessarily relativistic.
 Here, we assume relativistic movement
of the shock front. The magnetic field can be up to $B \sim 10^{-3}$~G in the jet at a radius 
of $r \sim 1$~kpc and the field typically decreases inversely with the radius. This condition 
allows for particle acceleration up to the highest energies, 
i.e.~$E_{\max}^{AGN}\sim 10^{21}$~eV \citep{BiermannStrit87}.
 Radio galaxy hot spots allow for acceleration of particles up to
$10^{21}$~eV and energies of $10^{19}$~eV can be
produced in radio galaxy lobes. Other suggested sources are not able to produce particles of 
sufficient energy because the field strengths available are insufficient to contain the particles. 

GRBs may also 
accelerate protons to $5\times10^{20}$~eV. This is because fields up to $\Gamma B=10^{14}$ G can be
produced at an accretion torus at about $6\times10^{6}$ cm radius
in massive star collapse before quantum mechanical dissipation, limits the field amplification
\citep{Lerche77} and the subsequent $1/r$ field fall-off yields an $E_{\max}$ independent of $r$. 
 Acceleration occurs without significant synchrotron losses,
see~\cite{vietri95,waxman_emax}. The acceleration of particles during the
prompt emission phase in highly-relativistic shocks ($\Gamma=100-1000$)
is discussed in the following. 
Acceleration in external shocks during the slowing afterglow phase
is also possible so that
 the spectra may resemble the AGN spectra that
we calculate and present in Section~\ref{results}.
\section{The physical concept and the Monte Carlo simulations\label{simulations}}
\cite{begelman_kirk90} have claimed that most upstream field
configurations at high shock boost factors $\Gamma$ appear \textit{superluminal}. 
In superluminal shocks the particles
are accelerated in a shock drift because there is no transformation into a de Hoffmann-Teller (HT) 
frame \citep{ht1950}, where $\vec{E}=\vec{0}$. 
To transform from the normal shock frame (NSH) to the HT frame, we 
need to boost by a speed $V_{HT}$ along the shock frame, where $V_{HT}= V_{NSH} \cdot \tan\psi$ 
($\psi$ is the angle between the magnetic field and the shock normal).
Due to physical causality, this transformation is only possible if $V_{HT}$ is less or 
equal to the speed of light. Thus, when $V_{NSH}=c$, the limit is $\tan \psi=1$. When $\tan\psi\leq 1$ 
the \textit{subluminal} 
shock transformation case is applied. For all other cases, where a HT frame cannot be found due to a very high
inclination in combination with a high shock velocity, the superluminal shock condition applies.
While in the subluminal case particle transmission at the shock can be decided in the HT frame employing
conservation of the first adiabatic invariant, in the superluminal case computations are followed
entirely in the fluid rest frames with reference to the shock frame simply employed 
to check whether upstream or downstream shock conditions apply. 

Superluminal shock acceleration may be treated as a
shock drift mechanism in the shock frame and is best visualised
when the shock is nearly perpendicular.
As viewed in the shock rest frame, the particle
is moving in a steep magnetic field gradient perpendicular to the shock surface.
The plasma motion at an angle to the magnetic
field creates an electric field given by $\vec{E}=-\vec{u}\times\vec{B}/c$. 
    In this mechanism, scattering is unimportant and particles can be accelerated by just one
shock encounter. However, due to the high field inclination, a particle
gyrating around the field lines crosses the shock more then once in one shock
encounter and thus obtains significant additional energy. 
The drift is in a direction perpendicular to the magnetic field gradient
(shock normal) and the magnetic field, according to
\begin{equation}
\vec{V}_d= \frac{p\cdot  c\cdot  u}{3e}(\frac{ \vec{B}\times\nabla B}{B^2})\,,
\end{equation}
with $B=|\vec{B}|$.
 The drift direction is such as to cause 
the particle's energy to increase. 
                   
  \cite{meli_quenby03_1,meli_quenby03_2} showed that a transformation from an initially
isotropic rest frame distribution to an accelerated flow frame leads to a comoving
relativistic plasma frame field distribution lying close to the flow vector.
This condition allows for a range of subluminal situations when viewed in the shock frame.
\cite{begelman_kirk90} pointed out, on the other hand, that in the blast wave frame the turbulence can be isotropic
and many shock stationary frame configurations can be superluminal.
In general, flow into and out of the shock discontinuity is not along the shock normal,  
but a transformation is possible into the NSH frame to render 
the flows along the normal ~\citep{begelman_kirk90}. We
assume that such a transformation has already been made.

\cite{vietri03} raise the question as to the correct reference frame in which jet acceleration 
should be viewed. To become CRs, the protons escape sideways out of the narrow beamed jet and are seen in the
extragalactic rest frame.  Escape upstream into the ambient medium is very unlikely. Strictly speaking,
we need to employ a transformation
conserving the distribution function $f$ depending on the momentum $p$, so $f(p)=f'(p')$ (\cite{Forman70} 
demonstrates this invariance which is related to Liouville's theorem). Here, $p^{2}f(p)=dJ(E(p),\vec{n})/dE$,
where $f(p)$ is in the direction of vector $\vec{n}$ and the energy $E$
corresponds to the momentum vector
$\vec{p}$. The particle flux $dJ/dE$ is typically in units of
cm$^{-2}$~ s$^{-1}$~sr$^{-1}$~GeV$^{-1}$. 
Prime and unprimed denote
rest frame and observer's frame.
For our calculations we choose the total energy normalisation to be done in the shock frame.
While the transformation affects the maximum energy obtained in the rest
frame to some extent, the fact that escape from the beamed jet is mainly by motion perpendicular 
to the flow, means that we are chiefly involved in transforming a momentum vector perpendicular 
to the relative velocity of the reference frames, where there is no Lorentz correction. 
For X-ray production in shocks the relevant results are in the downstream frame.

The purpose of the Monte Carlo simulations is to find a solution to the
particle transport equation for highly-relativistic flow
velocities. The appropriate time independent Boltzmann equation is given
by the following
\begin{equation}
\Gamma(V+\upsilon\mu )\frac{\partial f}{\partial x}=\frac{\partial f}{\partial t}\arrowvert_{c}\,,
\label{boltzmann:equ}
\end{equation}
where a steady state is assumed at the shock rest frame, and $V$ is the fluid velocity, $\upsilon$ the 
velocity of the particle, $\Gamma$ the Lorentz factor of the fluid frame, $\mu=\cos\theta$ the cosine 
of the particle's pitch angle $\theta$ and $\left.\partial f/\partial t\right|_{c}$
the collision operator. 
 
Particle scattering by magnetic irregularities fixed in the plasma frame will be assumed. The Alfv\'en velocity, $V_{A}$,
for AGN jets with a maximum $10^{-3}$ G field and a particle density of $10^{-2}$~cm$^{-3}$ is 
$2\times10^{9}$~cm/s. For GRB jets with a 10~G field and a particle
density of $10^{6}$~cm$^{-3}$, it is of similar magnitude. With these 
wave velocities, the waves will appear almost stationary to relativistic particles and second order Fermi acceleration, 
with a fractional energy gain $(V_{A}/c)^{2}$, is small compared with shock acceleration where a single encounter cause energy enhancement by a $\Gamma^{2}$ factor. $V_{A}\rightarrow c/\sqrt{3}$ is the theoretical relativistic limit to 
wave speed and to shock speed in the fluid frame while numerical jet simulation suggests large scale, knot features travel at
about c/10  \citep{van_Putten05}. Thus, we seem justified in neglecting fluid frame acceleration beyond the region
of trajectory intersection with the shock surface. This is especially true downstream where the second order Fermi 
fractional gain per collision is limited to 1/12. 

The scattering operator will be treated via a pitch angle scattering approach, rather than as previously 
in the work of \cite{meli_quenby03_1, meli_quenby03_2} where large angle scattering was considered. 
In standard kinetic theory the spatial diffusion 
coefficients $\kappa_{\|}$ and  $\kappa_{\perp}$, are related to the formula
$\kappa_{\perp}= \kappa_{\|}\cdot(1+(\lambda/r_l)^2)^{-1}$, \cite{jokipii87}. 
In the well known Bohm Limit, $\lambda/r_l =1$, but interplanetary particle propagation studies and gyroresonance
theory suggest $\lambda$ to be a number of particle gyroradii $r_l$, that is $\lambda\geq10 r_l$.
Hence, in the shock normal, or in the $x$ direction, 
the diffusion coefficient is given by $\kappa_{\|}=\lambda v/3$ where $\kappa=\kappa_{\|}\cos^{2} \psi$, 
since we assume that $\kappa_{\|}>> \kappa_{\perp}$. A guiding centre approximation is therefore used to
follow propagation along field lines.
Because relativistic shocks generate strong small-scale turbulent magnetic field downstream 
by the relativistic two stream instability \citep{Medvedev99}, we assume power in the scattering increases, as does the  
field strength and for simplicity we assume these quantities change to keep their ratio constant.
Therefore,  $\lambda_{down}= \lambda_{up}$.

\cite{Gallantetal99} have demonstrated analytically that particles entering the upstream region
in a direction nearly normal to the shock can only experience small-angle
scattering (pitch angle diffusion), $\delta \theta \leq 1/\Gamma$ with $\delta\theta$ measured 
in the upstream fluid frame for scattering in a uniform  field or a randomly oriented set of 
uniform field cells. The condition arises because particles attempting to penetrate upstream 
from the shock are swept back into the shock before they can scatter far from it.
If we consider the ratio of initial energy to final energy, measured in the upstream or 
'primed' frame, after up to down to up transmission in an inclined shock, we have
\be
E'_{f}/E'_{i}=\Gamma^2(1+\beta_{r}\mu'_{\rightarrow d})(1-\beta_{r}\mu*_{\rightarrow u})\,
\ee
where $\beta_{r}$ is the relative velocity of the frames as a fraction of c, $\mu'_{\rightarrow d}$
is the cosine of the particle velocity angle to the normal direction in the upstream frame at the moment
of going downstream and $\mu*_{\rightarrow u}$ is cosine of the particle velocity angle to the normal
in the downstream frame at the moment of going upstream. For down to up crossing, $-\mu*_{\rightarrow u}>B^t_{1}/B^t_{2}$
where $B^t_{1}/B^t_{2}$ is the inverse compression ratio.
Up to down, the pitch angle scattering constraint is $\mu'_{\rightarrow d}\approx -1+1/\Gamma^2$.
As pointed out by  \cite{Baring99}, $E'_{f}/E'_{i}\approx 2$. On the first up to down to up cycle
with injection upstream and directed towards the shock, $E'_{f}/E_{i}\approx\Gamma^{2}$, corresponding
to the large angle scattering case of  \cite{QuenbyLieu89}. 
Since this constraint is largely dependent on the 
kinematic competition between upstream particle flow and the relativistic approach of the shock front,
it is not critically dependent on the exact magnitude of the pitch angle scattering. Moreover, as shown 
by \cite{Quenby05}, blobs of high field scattering centres which could cause even large angle
scattering, are allowed if the Larmor radius in the blob, $r_{g,b}$, satisfies
$(\lambda/2\Gamma^{2})>r_{g,b}$ where the parallel mean free path $\lambda=Xr_{g,ambient}$ and $X\sim10$.
Such fields would allow for substantial scattering before particles are swept back to the shock. 
This criterion, derived for parallel shocks, effectively applies in the oblique, subluminal case
since in the upstream frame, the field direction is near the normal direction. To obtain a deflection
$\sim10\Gamma^{-1}$, the constraint is relaxed to $b/B>\approx2\Gamma$ as the ratio of 'blob'
to ambient field, where $b$ is the perpendicular perturbation to the mean field, $B$. 
Hence, a relatively few field blobs, perhaps originating in strong instability in a hypernova collapse 
and with field strengths up to a factor $1000$ stronger than the ambient, would thus be required to allow for
pitch angle scattering greater than $\delta \theta=1/\Gamma$. 
For practical purposes, remembering it is the downstream scattering that is relevant
to particle loss, it seems reasonable to use larger 
 values of $\delta\theta$,  within the pitch angle diffusion model.

 A simple representation for the effect of the turbulence which relates to previous work 
is to suppose the particle scatters $\delta \theta=N/\Gamma$ every $\lambda=10r_{l}$
where $ r_{l}$ is the Larmor radius, in the plane 
of gyration. A transverse field perturbation changes the pitch angle 
in a quasi-linear theory \citep{Kennel66}, by 
\be
\delta \theta =\omega\frac{b}{B}\delta t\,
\ee  
in a time $ \delta t$ due to a perpendicular perturbation, $b$, to the mean field,  $B$, with cyclotron angular frequency,
$\omega= eB/\gamma m_{\circ}c$. The particle moves in near gyroresonance with the wave in $b$.
A pitch angle diffusion coefficient can then be derived
\be
D_{\theta}=\frac{\delta\theta^{2}}{\delta t}=\frac{\omega^{2}}{v_{\|}}\frac{P(k)}{B^{2}} \,,
\ee 
where $P(k)=P_{\circ}k^{s}$ is the power spectral density of $b$ at gyroresonance wave number $k=\omega/v_{\|}$.
A particle then diffuses in pitch by a finite amount, $\delta \theta$ during $\delta t$  given by
\be
\delta \theta^{2}=2D_{\theta}\delta t \,.
\ee
It is waves, of wave number $k$, fulfilling the gyroresonance condition, $k=\omega/v_{\|}$ that cause the scattering
of particles satisfying $\omega r_{l}=v_{\perp}$. We choose $\delta \theta $ to lie between $1/\Gamma$ and $10/\Gamma$,
so on average a particle scatters $5/\Gamma$ after a time $10\sqrt{3}r_{l}/c$ so that
\be
\delta \theta^{2}=\frac{25}{\Gamma^{2}}=\frac{2\omega^{2}}{v_{\|}B^{2}}P_{\circ}(k)\frac{\omega^{s}}  {v_{\|}^{s}}
\frac{10\sqrt{3}v_{\perp}}{c\,\omega}\,.
\ee
 To choose $\delta \theta$ independent of particle $\gamma$ that is of
$\omega$, we require $s$=-1 for the spectral slope. Then we obtain a power spectrum relative to the mean field power
\be
\frac{P(k)}{B^{2}}=\frac{5}{4\sqrt{2}\Gamma^{2}}k^{-1}\,.
\ee
The total fractional power in the turbulence if resonating waves are present to scatter particles of between 
$\gamma=300$ and $\gamma=10^{12}$ is $1.4/\Gamma^{2}$. Hence, the chosen pitch angle scattering model corresponds to
a weak turbulence situation. Because quasi-linear theory for wave particle interactions is known to be an inexact
approximation, the power spectrum we have presented must also be an approximation to the scattering model we employ.
The choice of a fixed factor, 10, in the relation between $\lambda$ and $r_{l}$ acknowledges this inexactness.  
The spectrum is simply presented to provide a link with the work of others, especially \cite{Niemec05} who
realise the field fluctuations employing a specified wave spectrum.  
In the present investigation, we allow the
  particles with pitch angle chosen at random to lie in the range of  
$1/\Gamma \leq \delta \theta \leq 10/\Gamma$. In our past work mentioned in Section~\ref{introduction}, we presented results 
for the large angle case and the aforementioned extreme limiting case of $\delta \theta  \leq 1/\Gamma$.

Our past studies of pitch angle scattering \cite{meli_quenby03_1,meli_quenby03_2} suggested that
as the magnetic field inclination angle to the shock normal decreased, the spectra become smoother.  
Both a pronounced, plateau-like structure and increasing flatness developed for the highest values of the shock 
boost factor $\Gamma$. Additionally, for all inclination values used in the simulations, for $\Gamma = 10 - 30$, 
the spectral form remains smooth.
\cite{EllisonDouble04}, \cite{Steckeretal07} and references therein, have shown similar trends.
In this more comprehensive study, the previous claims will be more thoroughly investigated.

Standard theory poses the conservation of the first adiabatic invariant
in the HT frame in order to determine reflection or transmission 
of the particles. Since in this frame the allowed and forbidden angles for transmission
depend only on the input pitch and phase, not on rigidity, the results
of~\cite{hudson1965} apply in our model. For an isotropic flux, the transmission coefficient 
$\zeta$ is simply given by the
particle flux conservation between an upstream magnetic flux tube area and the corresponding
downstream flux tube area to which it connects
\be
\zeta= \frac{FB_{1}}{FB_{2}} \,,
\ee
where $F$ is the distribution function \citep{Parker65}. 
Trajectory integration, \cite{hudson1965} giving the phase dependence of the probability of transmission
as a function of phase and pitch angle showed that the reflection percent plotted against pitch angle 
 never varied more than $20 \%$ from the mean value and these results were consistent with
the flux conservation prediction based on \cite{Parker65} and the adiabatic invariant conservation. 
In the relativistic shock situation anisotropy renders the input to the shock from upstream, 
very anisotropic in pitch angle, but as was discussed in \cite{MeliThesis}, it is an
acceptable approximation to randomise phase before transforming to the HT frame
and then to use the adiabatic invariant to decide on reflection/transmission, because of 
the Hudson result.

For further details on the simulations, see appendix~\ref{simulation:appendix}.
\section{Results\label{results}}
The physical concepts and analytical approximations previously mentioned 
are used to perform Monte Carlo simulations 
for relativistic superluminal and subluminal shocks. Computed particle 
spectra will be presented with special focus on the relation between a given astrophysical source 
shock boost factor $\Gamma$ and the resulting spectral features.
Of particular relevance is Section~\ref{diffuse_crs:sec} devoted to the
calculation of the contribution of these sources to the observed diffuse CR spectrum. 
\subsection{CR shock acceleration spectra}

\subsubsection{Superluminal shock spectra}

Initially, we present simulations for relativistic superluminal shocks as described in the previous section and the Appendix.
Unlike our previous work which was confined to large angle scattering, pitch angle scatter is employed.
The guiding centre approximation is used except within one complete particle helical cycle from the shock. 
Since a transformation into the HT frame is not possible as described in Section~\ref{simulations}, the particles are
followed in the appropriate fluid rest and SH frames, simulating 
the physical picture of the shock drift mechanism. 
We follow the helix trajectory of the particle until it intersects
with the shock front, applying a pitch angle 
scatter [$\delta \theta \leq 10/\Gamma , \phi \in  (0, 2\pi)]$ right up to the shock interface.
Simulation runs performed showed that the results were basically independent of $\psi$, magnetic field
to shock normal angle.  We therefore show here a simulation run with an arbitrarily
chosen inclination angle, $\psi=76\deg$, as an example, employing a range of boost factors. 
The resulting spectra are presented in the shock
frame on the downstream side as we want to ensure comparability  
with following calculations.

\begin{figure}[ht]
\begin{center}
\epsfig{file=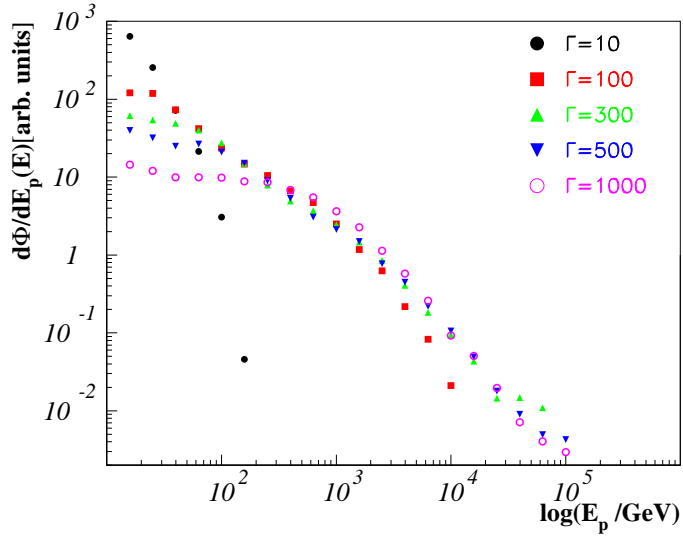, width=10cm}
\caption[Superluminal, relativistic spectra at $\psi=76\deg$]{Superluminal, 
relativistic spectra at $\psi=76\deg$. Boost factors are varied between 
$\Gamma=10,100,300,500,1000$. Spectra for different inclination angles $\psi$ are comparable.}
\label{super_spectra:fig}
\end{center}
\end{figure}

The resulting particle spectra for $\Gamma= 10, 100, 300, 500$~and~$1000$ are
displayed in Fig.~\ref{super_spectra:fig}. These plots indicate that for mildly-relativistic shocks
($\Gamma=10$), the acceleration is efficient up to energies
$\ep<10^{2}$~GeV. More highly-relativistic shocks ($\Gamma\geq 100$) can produce particles 
up to $\ep<10^5$~GeV. The upper limit of significant acceleration is $\sim \Gamma^{2}$, corresponding
to only one complete particle crossing cycle with the majority of particles either failing to return a second time
or only returning at angles close to the normal.
In the region of efficient acceleration, the spectra approximately follow
power-laws with spectral indices lying between $\sim 2.0 - 2.3$. In contrast, for the case of large angle 
scattering, previously studied by \cite{meli_quenby03_2}, the spectra could not be described by power-laws,
but exhibited a concave shape terminating in a steep energy cut-off.  

We conclude that superluminal, relativistic shocks are not efficient accelerators 
for very high energy particles and are unlikely to contribute to
observable effects, discussed in more detail in
Section~\ref{diffuse_crs:sec}. These conclusions concur with the work
of \cite{niemec_ostrowski_icrc07}. While superluminal shocks cannot
contribute to the observed spectrum of charged UHECRs, a contribution
to the neutrino- and TeV-photon background arising from proton-photon
or proton-proton interactions, is still possible as discussed in Section~\ref{neutrinos:sec}.
\subsubsection{Subluminal shock spectra}
Particle spectra produced in relativistic subluminal shocks have been
calculated for three different inclination angles in the shock frame,
$\psi=23^{\circ},\,33^{\circ}$~and \\$43^{\circ}$ which we chose as
representing the possible range of subluminal shock angles likely in astrophysical sources.
In appendix~\ref{spectra:appendix}, simulated particle spectra are
presented for the entire energy range of particle energy considered, $E=10^{2}- 10^{12}$~GeV, for
a range of boost factors, $\Gamma=10$ to $\Gamma=1000$ and for all three angles.
The chief result is that the spectra appear as  smooth power-laws for mildly-relativistic shocks,
$\Gamma \leq 30$. As the boost factor increases, the spectra appear with bumps
($\Gamma>100$): The plateau-like parts of the spectra at higher energies and higher boost
factors are caused by particles continuing to undergo significant acceleration in a second cycle. The lower energy part
of the spectrum is dominated by particles undergoing one acceleration shock crossing
while the second bump in the spectrum represents particles experiencing two acceleration shock crossings.
In this first complete shock cycle crossing from upstream to downstream to upstream, the energy gain is a factor
$\Gamma^{2}$, operating an injection energy already $\approx\Gamma$ in magnitude. Subsequent crossings become smoother
since the energy gain is expected to be limited to $\sim2$ and there is a statistical smoothing of the energy gains.
At the highest boost factors we investigate, this smoothing regime is not reached.
    
\begin{figure}
\centering{
\epsfig{file=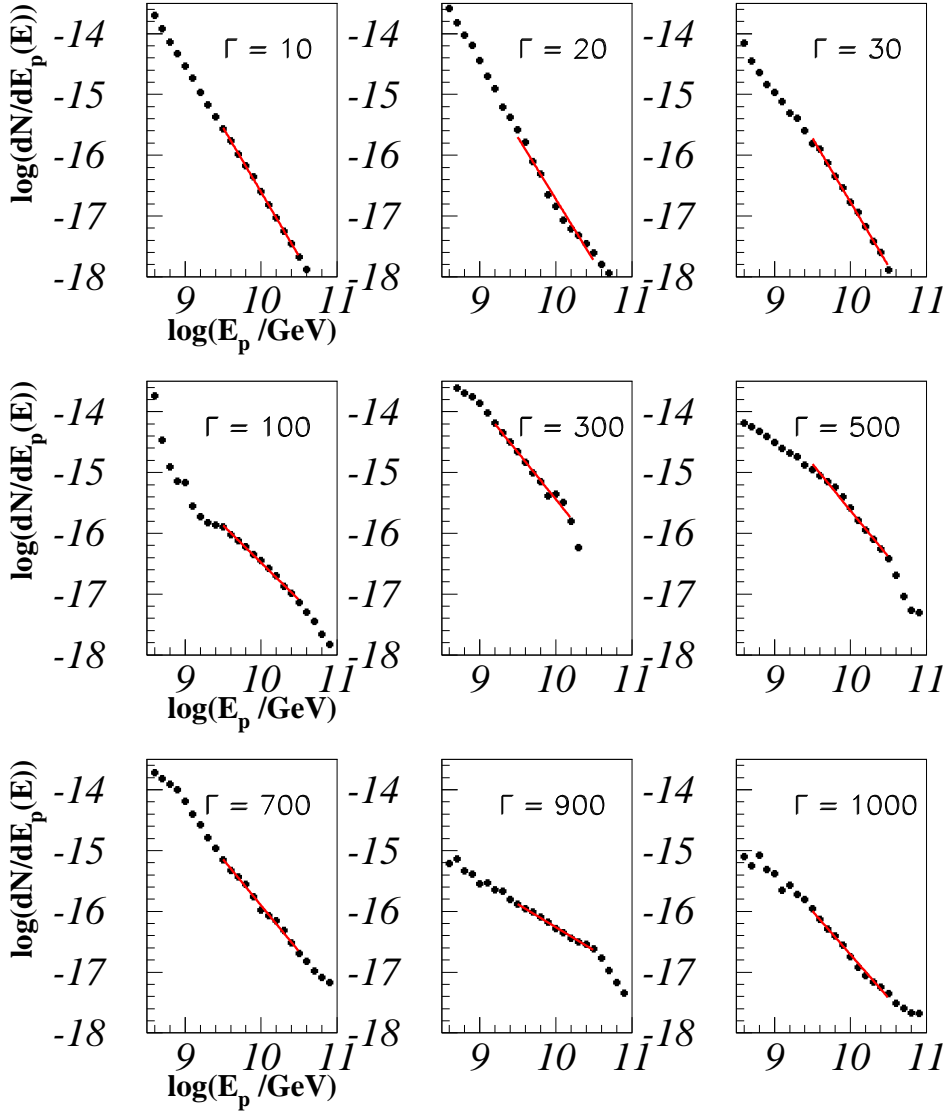,width=\linewidth}
\caption{Subluminal spectra averaged over the three angles for different
  $\Gamma$: $\Gamma=10,20,30$ is displayed in the first row, in the middle,
  $\Gamma=100,300,500$ is shown and $\Gamma=700,900,1000$ is the bottom
  row. The black crosses in each graph represent the simulation result. The straight
  line shows the single power-law for comparison.  \label{mean_all}}
}
\end{figure}

In order to get a representative picture, we average the particle spectra over the
three angles at a particular $\Gamma$, in all following calculations. This should give a more realistic view
of the diffuse particle flux from extragalactic sources, as it is expected
that a range of angles is likely to occur in the class of AGN and GRB
shocks. 
We concentrate on 
the highest energies because observation of particle-induced air showers at
energies between $10^{9.5}$~GeV and $10^{10.5}$~GeV indicates an extragalactic
origin of the charged CRs, distinct from a dominant, galaxy produced component at lower energies. A
power-law fit is made to the simulated spectra between $10^{9.5}$~GeV and
$10^{10.5}$~GeV. Figure~\ref{mean_all} shows the averaged, simulated spectra
between $10^{8.5}$~GeV and $10^{11}$~GeV. 
 At even higher energies, the spectrum is altered
by the absorption of protons due to interactions with the cosmic microwave
background. While the normalisation of the spectra is arbitrary since
dependent on the number of injected particles in the Monte-Carlo simulation,
the spectral index can be compared to what is observed in CRs in the
same energy range. Table~\ref{gamma_index_mean:tab} shows the variation of the
spectral index with the boost factor. While mildly-relativistic shocks show
indices around $\alpha_p\approx 2$,  highly-relativistic shocks with
$\Gamma>100$ have flatter spectra between $0.7<\alpha_p<1.5$. Particle
spectra emitted during the prompt phase of GRBs ($\Gamma>100$) will therefore appear much flatter than
AGN particle spectra ($\Gamma\sim 10$). This has important implications for the
interpretation of the origin of the UHECR spectrum as will be discussed in a
following subsection.

\cite{Steckeretal07} investigating parallel shocks up to $\Gamma=30$ found increasing spectral structure and decreasing slope as $\Gamma$ increased ($E^{-1.26}$ at $\Gamma=30$),
trends we have shown to extend to far higher $\Gamma$ factors and for a more general class of
subluminal shock inclination angles. \cite{bednarz_ostrowski98} used pitch angle scattering and varying cross-field diffusion
and found that at low $\Gamma$, steep spectra occurred at large inclination angles but all values of
these parameters seemed to produce spectral slopes of -2.2 at $ \Gamma=243$. In contrast, \cite{meli_quenby03_1}
found spectra flatter than $E^{-1}$ for parallel shocks as $\Gamma\rightarrow1000$.
Later work, however, with wave spectra 
$P(k)\sim k^{-1\rightarrow1.5}$ by \cite{Niemec05} found spectra flatter than $E^{-1}$ with 
noticeable structure spectra in inclined, subluminal shocks at upstream velocities of 0.5c. Their trajectory integrations 
took into account cross field diffusion. \cite{baring04b} however, cites previous work with no cross field diffusion
which finds significant relativistic, inclined shock acceleration to be limited to inclination angles $\leq25^{\circ}$.
It is not clear whether the very steep spectra quoted at higher inclination angles are due to the steep slope 
at the edge of the first plateau we mention above, or whether there is a significant difference from our modelling.
A high perpendicular diffusion coefficient, high inclination shock situation might be expected to approach a high scattering,
parallel shock regime, but this possibility does not seem to reconcile the various conflicting results just discussed.
There is an agreement that at very high inclinations significant acceleration above 
that due to a single shock cycle is ruled out. Also, there seems to be a developing consensus that high $\Gamma$,
subluminal shocks result in flatter spectra than $E^{-2}$. 
\cite{Dingus95} provides gamma-ray burst evidence for relativistic
electron spectra with relatively flat
slopes with exponents at least as low as $\approx-2$.   

Results indicating a variety of possible slopes are also consistent with radio data on
the electron spectra injected at terminal hotspots in the lobes of  
powerful FR-II radio galaxies where no single, universal power-law is found, 
as shown by \cite{Rudnicketal94} and \cite{Machalskietal07} among others.

\begin{table}
\centering{
\begin{tabular}{|l||l|l|l|l|l|l|l|l|l|}
\hline
$\Gamma$  &10 &20 &30 &100&300&500&700&900&1000\\\hline
$\alpha_p$&$2.1$&$2.0$&$2.1$&$1.2$&$1.5$&$1.5$&$1.5$&$0.7$&$1.4$ \\\hline
\end{tabular}
\caption{Spectral indices for a single power-law comparison for subluminal
 shocks. The spectral fits were made between $10^{9.5}$~GeV and $10^{10.5}$~GeV in
 order to be comparable to the observed CR flux at the same
 energies. The uncertainty from the fit is less than 10\% if we assume an
 accuracy of the simulation is better than $\Delta \log(dN/d\ep)\sim 0.5$, which is a
 conservative estimate. \label{gamma_index_mean:tab}}
}
\end{table}
\subsection{Diffuse CR spectra from GRBs and AGN\label{diffuse_crs:sec}}
The source spectra derived previously can be translated into an
expected diffuse proton flux from astrophysical sources by folding
the spectra with the spatial distribution of the sources. 
In this section, AGN and GRBs are
used as potential candidates because these 
are the sources with the highest observed output in relativistic electrons.
Since the
particle spectra are strongly dependent on $\Gamma$, it is
important to discuss which spectra to use for these two source classes. 
Spectral choice is investigated in the next subsection 
before the actual calculation of the diffuse spectra is shown. In the last
subsection, the results of our calculations are compared to CR data.
\subsubsection{AGN and GRBs - intrinsic spectra}
Spectral fits limited to the energy range $10^{9.5}$ GeV to $10^{10.5}$ GeV of extra-galactic origin
are employed for both AGN and GRB sources, but the boost factors applicable differ between these source classes.  

The boost factor deduced from electron synchrotron observation 
can vary significantly in the case of GRBs. While the
majority of sources are estimated to have boost factors around $\Gamma\approx
300$, more moderate values down to $\Gamma=100$ or more extreme values up to
$\Gamma=1000$ are believed to occur. However, the exact distribution of Gamma Ray
Burst $\Gamma$ factors cannot be determined. In many cases, only upper
limits can be given. In addition, there may be hidden bursts not observed
with GRB satellite experiments. It is therefore not useful to model a detailed distribution of
boost factors for GRBs, while the simulation results connecting the
spectral index of the spectrum with a boost factor are subject to uncertainty, as is implied by the absence of a 
monotonic trend in Table~\ref{gamma_index_mean:tab}.

 It was shown that for boost factors of $\Gamma>100$, the source
spectra lie between ${\ep}^{-1.5}$ and ${\ep}^{-0.7}$. A conservative estimate for the flattened GRB
spectra will be adopted using
\begin{equation}
\frac{d\Phi_{GRB}}{d\ep}\propto {\ep}^{-1.5}\,.
\end{equation}

 The situation is simpler for AGN, as the shock's boost factors vary only up to
between $\Gamma=10$ and $\Gamma=30$~\citep{BiermannStrit87}. Here, all computed spectra cluster around a value of
${\ep}^{-2.1}$. Therefore, the AGN spectrum will be taken as
\be
\frac{d\Phi_{AGN}}{d\ep}\propto {\ep}^{-2.1}\,.
\ee
\subsubsection{From CR shock acceleration spectra to a diffuse spectrum\label{spectratodiffuse}}
The diffuse spectrum as measured at Earth depends on several factors:
\begin{itemize}
\item {\it Single source spectra at the source} $d\Phi/d\ep$. The spectral
  behaviour was already discussed in the previous subsection. To account for
  particle propagation, adiabatic energy losses need
  to be considered as $\ep(z)=\ep\cdot (1+z)$. Here, $\ep(z)$ is the energy as observed at
  a source at redshift $z$ and $\ep$ is the corresponding energy observed at
  Earth. Diffusive propagation in the magnetic field between clusters is assumed to involve
 only small angle scattering with preservation of spectral shape. Anisotropy in the source distribution is neglected. 
 In addition, we consider pure proton spectra so that
  spallation effects are not present.
\item {\it Source evolution} $g(z)$: It is assumed that both AGN and GRBs
  follow the SFR to determine the
number density evolution with comoving volume, see for
  example, \cite{hasinger05} in the case of AGN and \cite{pugliese00} in the
  case of GRBs. A large sample of radio quiet
AGN selected at X-ray wavelengths was investigated by \cite{hasinger05}. The comoving density $dn/dV(z)$ is given as

\be
\frac{dn}{dV}(z)\propto 
\left\{ 
\begin{array}{lll}
(1+z)^{m}&&\mbox{for }z<z_1\\
(1+z_1)^m&&\mbox{for }z_1<z<z_2\\
(1+z_1)^m\cdot 10^{k\cdot (z-z_2)}&&\mbox{for }z>z_2\,, 
\end{array}
\right.
\ee
with the parameters $m=5.0$, $z_1=1.7$, $z_2=2.7$ and $k=-0.43$.
The total redshift evolution $g(z)$ further includes multiplying the comoving volume
$dV/dz$ with a factor $1/(4\,\pi\,d_{L}^{2})$ to account for the decrease of
the flux $L$ with the distance $d_L$, neglecting a possible travel limitation to the 
distance reached by significant magnetic scattering. Therefore,
\be
g(z)=\frac{dn}{dV}(z)\cdot \frac{dV}{dz}\cdot (4\,\pi\,d_{L}^{2})^{-1}\,.
\ee
For simplicity, this model is used
for both AGN and GRBs. Although deviations between the SFR scenarios of AGN and
GRBs are expected, the approximation that both follow the distribution of
radio quiet X-ray AGN is reasonable \citep{hasinger05}: the deviations being expected to be
negligible with respect to general uncertainties arising from assumptions about the acceleration region.
\item {\it Absorption of protons at the highest energies}: Protons at
  $\ep >5\cdot 10^{19}$~eV are absorbed due to interactions with the cosmic microwave
  background as was recently confirmed by the Auger experiment~\citep{yamamoto_icrc2007}.
 Therefore, the diffuse spectrum resulting from the
  propagation of a single source spectra is modified by a further factor,
  $\exp[-\ep/(5\cdot 10^{19}$~eV$)]$ to account for this effect.  
\item {\it The normalisation of the diffuse spectrum}: Because the calculated
  particle spectra are given in arbitrary units, normalisation of the
  overall spectrum as measured at Earth is achieved using observation. 
\begin{itemize}
\item[-] In the case of superluminal sources,  normalisation of the expected signal follows from the most
restrictive upper limit on the neutrino signal from extraterrestrial sources
given by the AMANDA experiment, see \citep{jess_diffuse}
\be
E_{\nu}^{2}\,\frac{dN_{\nu}}{dE_{\nu}}<7.4\cdot 10^{-8}\frac{\mbox{GeV}}{\mbox{s sr cm}^2}\,.
\ee
With an average $E^{-2}$ spectrum for both neutrinos and protons, the spectra are
connected by assuming that the expected neutrino energy fluence is a fraction
$q$ of the proton spectrum
\be
\int \frac{dN_{\nu}}{dE_{\nu}}\,E_{\nu}\,dE_{\nu}=q\cdot\int \frac{dN_p}{dE_p}\,E_p\,dE_p \,,
\ee
with $q=1/40$, since only $20\%$ of the proton flux goes into pion production
via the delta resonance, $1/2$ of the remaining flux goes into the charged
pion component of which $1/4$ goes into neutrinos, see e.g.~\cite{julia_review}.
\item[-] In the case of subluminal sources, using neutrino flux limits leads
to an excess above the observed spectrum of charged CRs, since the limits are
  not stringent enough yet. Instead, the measured
part above the 'ankle' of the CR spectrum is used for an estimate of the
contribution from subluminal sources. The CR energy flux above the
ankle is given by~\citep{wb97,wb99}
\be
j_E(E_{\min}=3\cdot 10^{18}\mbox{ eV}):=\int_{3\cdot 10^{18}\mbox{eV}}\frac{dN_p}{d\ep}\,\ep\,d\ep \\[0.2cm]
\approx 10^{-7}\,\diffunits\,.
\ee
It is expected that this contribution comes from a combined signal from AGN
and GRBs. In the following it is assumed that the fraction of UHECRs coming
from AGN, contributes a fraction $0<x<1$. 
 Therefore, the fraction of UHECRs from GRBs is $(1-x)$.
\end{itemize}
\end{itemize}
Thus, the total spectrum as observed at Earth is given as
\begin{small}
\be
\frac{dN_p}{d\ep}=A_p\int_{z_{\min}}^{z_{\max}} \left( x\cdot
\frac{d\Phi_{AGN}}{d\ep}(\ep(z))+(1-x)\cdot \frac{d\Phi_{GRB}}{d\ep}(\ep(z))\right)\cdot g(z)\,dz\,.
\ee
\end{small}
The minimum redshift is set to $z=0.0018$ as the distance of the closest AGN,
Centaurus~A. The maximum redshift is taken to be
$z_{\max}=7$. As the main contribution comes from redshifts of $z\sim 1-2$ due
to the high number of sources at these redshifts, the exact values of the
integration limits are not crucial.

\subsubsection{Comparison with the observed cosmic ray spectrum}

\begin{figure}
\begin{center}
\epsfig{file=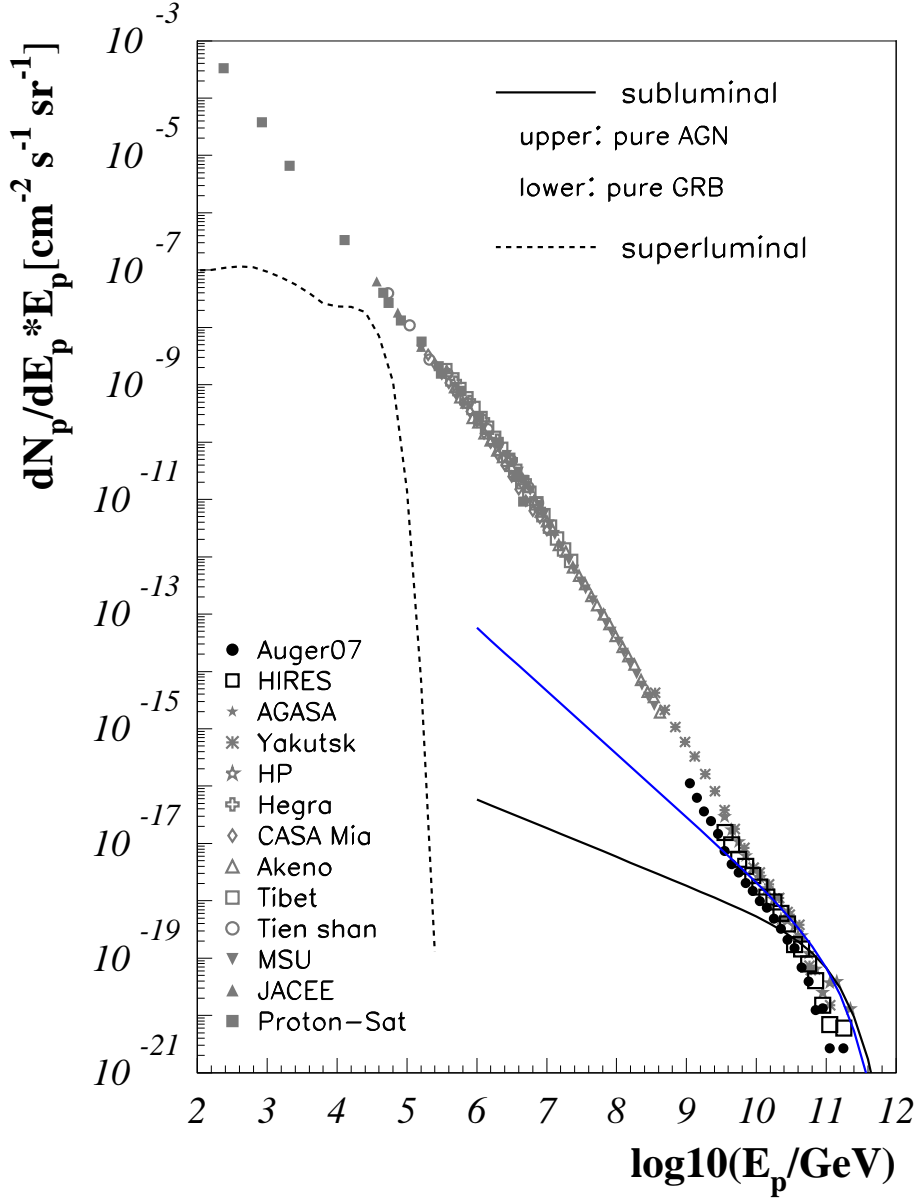, width=\linewidth}
\caption[Diffuse primary flux from GRBs and AGN]{The maximum predicted diffuse flux from GRBs and AGN with superluminal shock fronts (dashed line) and subluminal shocks (solid lines). For subluminal
sources, the upper line is a pure AGN-produced spectrum, the lower line represents 
a pure GRB spectrum. The flux is compared to the measured CR
spectrum. Data points are taken from the different experiments: Auger --
\cite{yamamoto_icrc2007}; HiRes -- \cite{hires_spect2002}; AGASA --
\cite{agasa_spect1995}; Yakutsk -- \cite{yakutsk_spect1985}; Haverah Park --
\cite{haverah_park_spect2001}; HEGRA -- \cite{hegra_spect1999}; CASA-MIA --
\cite{casa_mia_spect1999}; Akeno -- \cite{yakutsk_spect1985}; Tibet --
\cite{tibet_spect2003}; Tien Shan -- \cite{tien_shan_spect1995}; MSU --
\cite{msu_spect1994}; JACEE -- \cite{jacee_spect1995}; Proton-Sat --
\cite{proton_sat_spect1975}; KASCADE -- \cite{kascade_spect2005}. In the case of superluminal sources, 50\% is assumed
 to come each from GRB and from AGN.}
\label{sl_cr}
\end{center}
\end{figure}

The diffuse spectrum as measured at Earth is shown in Figure~\ref{sl_cr}. Data
points represent measurements from a selection of experiments. 
Our calculated spectra from superluminal and subluminal shocks are displayed 
as the dashed and solid lines. 

\begin{figure}
\begin{center}
\epsfig{file=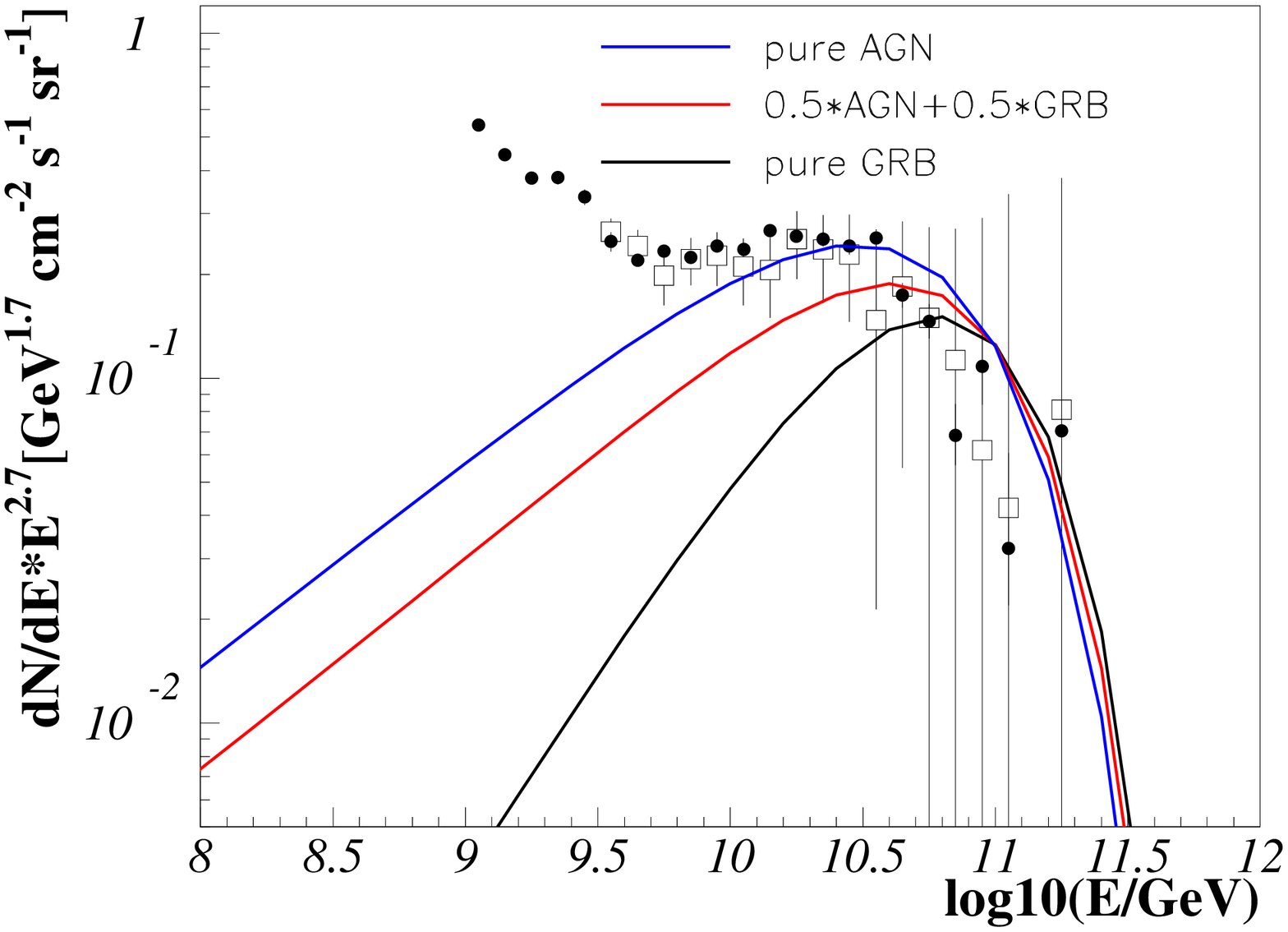, width=\linewidth}
\caption[Spectrum of UHECRs multiplied by $E^{2.7}$.]{Spectrum of UHECRs
  multiplied by $E^{2.7}$. Data points from Auger
\citep{yamamoto_icrc2007} and HiRes \citep{hires_spect2002}. The solid lines
represent the same predictions as presented in Fig.~\ref{sl_cr}. Auger data
have been renormalised at $10^{10.3}$~GeV to HiRes data and the calculated
spectra have also been normalised to HiRes data. The data can
be described well by a pure AGN spectrum (blue line) within experimental
uncertainties and test particle acceleration accuracy. The red line is a
mixture of 50\% GRB contribution and 50\% AGN contents, the black line is a
pure GRB spectrum. Both do not fit the data.}
\label{sl_cr_e27}
\end{center}
\end{figure}

It appears from Figure~\ref{sl_cr}, that in the superluminal shock case, the only possible 
contribution to the measured CR spectrum is around the knee. It is expected, however, that the effective 
flux is actually even lower, because the normalisation is based upon the assumption that the contribution 
cannot be more than the current neutrino flux limits permit. Therefore, the calculated
flux can be considered as an absolute upper limit. Here, the fraction of AGN protons has been chosen 
to be $x$=0.5, assuming that 50\% of the signal is produced by AGN and 50\% by GRB

The subluminal shock case has been investigated for different scenarios. The upper
line represents a spectrum that would be produced by AGN only
$(x=1)$. The lower line represents a pure GRB
spectrum ($x=0$). The flux is too low to explain the observed component above
the ankle if a significant contribution comes from GRBs because of the flatness of the
GRB spectra. It seems that  the flat spectra do 
not fit the present observations. 
Thus, AGN are the favoured sources for the production of UHECRs.
Figure~\ref{sl_cr_e27} shows the CR spectrum at the highest energies,
multiplied by $E^{2.7}$ in order to have a clearer view on the features of
the spectrum. The upper and lower solid lines represent the same predictions for 
shocks as in Fig.~\ref{sl_cr}.
The middle line assumes that $50\%$ is made up by AGN and the
remaining $50\%$ comes from GRBs
 The normalisation fits the HiRes data. There is
a discrepancy between the normalisation of HiRes and Auger data, which is not
entirely understood yet, but is probably due to systematic errors in the
energy and flux determination of the experiments. Therefore, the Auger data
are renormalised at $10^{10.3}$~GeV to match the normalisation of the HiRes
data.
With an assumed significant contribution 
from GRBs, it seems difficult to  explain the observed spectrum.
Within the uncertainties of the experimental data, our predicted spectra fit the measured spectra between $10^{9.8}$ GeV
and $10^{10.5}$ GeV if pure AGN spectra are assumed. The flux is 
slightly too low for the first data points above the ankle, between
$10^{9.5}$~GeV and $10^{9.8}$~GeV. The steepening of the spectrum at
lower energies cannot be due to some contribution from sources with a
high field inclination to the normal because we have shown that such
superluminal shocks cannot produce energies above about
$10^{10.5}$~GeV. It is possible that the steepening arises from the
addition of subluminal sources with varying maximum energy, determined
by varying maximum field strengths, so only a fraction of the sources
reach the highest energies of $10^{21}$~eV.~\cite{ahlers2005}, fitting
HiRes data to a cosmological source distribution similar to ours in
Section~\ref{spectratodiffuse}, find some steepening around
$10^{9}$~GeV due to details of the photo-production propagation
function. While these
effects suffice to explain the small difference for the lowest
energies above the ankle concerning AGN spectra, they cannot explain
the large differences between the data and the GRB spectral predictions.
\section{Implications for high energy neutrino and photon astronomy\label{neutrinos:sec}}
Fig.~\ref{sl_cr} has an interesting implication for neutrino and
TeV-photon astronomy. Neutrinos and TeV photons are produced in
proton-photon or proton-nucleon interactions, see
e.g.~\cite{julia_review} for a review,
\begin{eqnarray}
p\, \gamma &\longrightarrow&
\Delta^{+}\longrightarrow
\left\{
\begin{array}{lll}
p\,\pi^{0} &&\mbox{, fraction }2/3 \\ n\,  \pi^{+}&&\mbox{, fraction }1/3
\end{array}\right. \\
p\, p &\longrightarrow&\pi^{+}\,\pi^{-}\,\pi^{0}\,.
\end{eqnarray}
The decay of the $\pi^0$ leads to high energy photon emission, and
$\pi^{\pm}-$ particles produce neutrinos. The photon signal at TeV
energies is not unique, since leptonic processes like Inverse Compton
scattering contribute at the same energies. Therefore, the best,
unambiguous way
of identifying the hadronic interactions are neutrino observations.

Models for neutrino emission in AGN are present in
e.g.~\cite{mannheimjet,mpr,stecker2005,bbr_05}, where it is assumed that protons
accelerated in AGN jets can interact with different photon
fields to produce neutrinos. 

Proton-photon interactions in the prompt phase of GRBs can lead to
neutrino and TeV-photon production. In the first
approach by \cite{wb97,wb99}, it is assumed that GRBs are the sources
of UHECRs to calculate the neutrino spectrum. According to our results
in the previous sections, GRBs are unlikely to be the sources of UHECRs and the model does not hold anymore. However,
further developments of the model normalize the flux to the
electromagnetic output rather than to the flux of UHECRs, see
e.g.~\cite{guetta2004,bshr2006,murase_nagataki2006}. Here, the protons do not
need to be accelerated to the highest energies: The photon field of
the prompt emission of GRBs has characteristic energies of around
$\sim 100$~keV$- 1$~MeV, and can reach energies up to $>100$~MeV, see e.g.~\citep{egret_grbs1,egret_grbs2,egret_grbs3}. The proton energy necessary to produce a Delta
resonance, and with this TeV-photons and neutrinos, is given as
\be
E_p\geq
\frac{\Gamma^2}{(1+z)^2}\frac{m_{\Delta}^{2}-m_{p}^{2}}{4}\cdot
E_{\gamma}\,.
\label{ep_egamma:equ}
\ee
Here, $m_{\Delta}$ is the mass of the Delta resonance, $m_p$ is the
proton's mass, $z$ is the redshift of the source and $E_{\gamma}$ is
the characteristic photon energy. The proton energy required for the
process is therefore given as
\be
E_p\geq 5\cdot 10^{7}\cdot (1+z)^{-2}\cdot
\left(\frac{\Gamma}{100}\right)\cdot \left(\frac{E_{\gamma}}{1\,\rm{MeV}}\right)^{-1}\,\rm{GeV}\,.
\ee
Thus, with redshifts typically of the
order of $z=1$ and boost factors of around $\Gamma=300$, the proton
energy sufficient for neutrino and TeV-photon production is as
low as
\be
E_p\sim 10^{6}\,{\rm GeV}\,. 
\ee
This opens the possibility of neutrino production from sources
which are {\em not} observable in charged UHECRs. While the lower energy spectra of
extragalactic, charged cosmic rays cannot be observed due to the high
galactic background, neutrinos may serve to investigate those sources
further.

Using the same reasoning as above, we can conclude that charged UHECRs
from {\em superluminal}
spectra may not be observed in charged cosmic rays, but are good
candidates for the production of high energy neutrinos and photons
from AGN or GRBs. Those shocks may produce low energy spectra of much higher
intensity than subluminal shocks, which are simply hidden due to the
galactic background. Neutrinos and photons, on the other hand, point
back to the original source, and may be identified.  We can use the maximum energy of the proton spectra
to calculate the energy of the photon spectra which are necessary to
produce high energy photons and neutrinos by using
Equ.~(\ref{ep_egamma:equ}),
\begin{eqnarray}
E_{\gamma}&\geq& \frac{\Gamma^2}{(1+z)^2}\frac{m_{\Delta}^{2}-m_{p}^2{2}}{4}\cdot
E_{p}\\
&=& 5\cdot 10^{7}\cdot (1+z)^{-2}\cdot
\left(\frac{\Gamma}{100}\right)\cdot \left(\frac{E_{p}}{1\,\rm{GeV}}\right)^{-1}\,\rm{MeV}\,.
\end{eqnarray}

For mildly relativistic shocks of $\Gamma\sim 10$ as they occur in
AGN, superluminal proton spectra reach up to $E_p \sim
100$~GeV. This requires photon energies of $E_{\gamma}\geq
10$~GeV. These high energies can be produced by Inverse Compton
scattering of synchrotron or external photons with the accelerated
electrons. A catalog of AGN with photon emission above $100$~MeV was
already presented by the EGRET experiment~\citep{egret}, and
more sources are likely to be identified when GLAST is launched this
summer \citep{glast}.

Highly relativistic shocks with $\Gamma>100$ reach up to proton energies of $10^{5}$~GeV. This
requires photon energies of $E_{\gamma}\geq 100$~MeV. This is about 2
orders of magnitude above the characteristic energy, but the photon
spectrum is likely to extend to energies above $100$~MeV as already observed
for more than 30 GRBs, see
e.g.~\citep{egret_grbs1,egret_grbs2,egret_grbs3}. If mildly
relativistic internal GRB shocks are to contribute to neutrino
production, photon energies in excess of $1$~GeV are required. Such
sources would also need to be insignificant producers of UHECRs, to
satisfy the discussion of Section~\ref{spectratodiffuse}.

In conclusion, low energy proton spectra from extragalactic sources
cannot be observed directly due to the high galactic background of
cosmic rays, but they may be detected indirectly by means of high
energy neutrino and photon spectra.
\section{Summary \& conclusions\label{summary}}
In this work we have presented Monte Carlo simulation studies of the acceleration of test
particles in relativistic, subluminal and superluminal shock environments. The source
candidates discussed were AGN jets with mildly-relativistic shocks of 
boost factors of $\Gamma\approx 10-30$ and GRBs regions with highly-relativistic shocks,
$100<\Gamma<1000$. The resulting particle spectra were used to calculate a contribution
to the diffuse CR spectrum.

Particle spectra have been obtained
with varying the shock boost factor $\Gamma$ and shock obliquity, 
i.e. the inclination angle between the shock normal and the magnetic field, $\psi$. 
Only subluminal shocks are efficient enough to accelerate particles 
up to $10^{12}$~GeV, while superluminal shocks are
effective up to $\sim 10^{5}$~GeV. 
Flat spectra are found for very high subluminal shock boost factors, but for superluminal shocks the spectral indices
stay roughly constant between values of 2.0 to 2.3 in the limited region of efficient acceleration, before a cutoff sets in.
 For the subluminal shock cases, 
the spectra for mildly-relativistic shocks have spectral indices around
$2.0-2.2$. Highly-relativistic shocks have spectra as flat as
${\ep}^{-0.7}-{\ep}^{-1.5}$ at energies between $10^{9.5}$~GeV and $10^{10.5}$~GeV.
There is no universal spectral form, rather a variety of spectral shapes with a noticable plateau-like structure developing
at higher $\Gamma$ values. This structure is very probably related to the number of scattering cycles undergone by particles
at a particular energy.
 
Our results can be summarised as follows:

\begin{enumerate}
\item Subluminal shock studies with the pitch scattering angle determined to lie in the range
 $1/\Gamma \leq\delta\theta\leq10/\Gamma$, approximately correspond to a situation with a spectrum of scattering waves,
$P(k)\,B^{2}=5/4/\sqrt{2} \cdot \Gamma^{-2}\cdot k^{-1}$ and with the neglect of cross-field diffusion.
The resulting spectral slopes were roughly independent of inclination angle, though details of the features were different. 
A dependence of the spectral index $\alpha_p$ on the shock boost factor $\Gamma$ 
was found, leading to spectra of $\alpha_p\sim 2.0 - 2.1$ for mildly-relativistic shocks of
$\Gamma\sim 10-30$, but producing much harder spectra ($0.7<\alpha_p<1.5$) for highly-relativistic
shocks, $100<\Gamma<1000$.

The above implies that GRB particle spectra arising from relativistic shocks
with very high boost factors between $100<\Gamma<1000$, have spectra flatter than
${\ep}^{-1.5}$, much flatter than AGN spectra, $\sim {\ep}^{-2}$. 
Moreover, the above findings are supported by the lower $\Gamma$ work of \cite{Niemec05} and
\cite{Steckeretal07}. Observational evidence,
\citep{Dingus95}, regarding irregular and flat spectra from GRBs may be explained by the spectra we present. 
This work is also consistent with the general observations of 
the electron spectra that may be injected from the terminal hotspots to the lobes of the 
powerful FR-II radio galaxies which are not of a single and universal power-law form, 
as shown in detail in \cite{Rudnicketal94}, \cite{Machalskietal07}, etc. 

\item Superluminal shocks are only efficient in accelerating CRs 
up to $\ep\sim 10^{5}$~GeV, resulting in spectral indices of  $\alpha_p\sim 2.0-2.3$. On the other hand,
subluminal shocks are more efficient and able to accelerate CRs up to $\ep\sim 10^{12}$~GeV, factors
of $10^{9\rightarrow11}$ above the particle injection energy. 

\item We discussed the possible contributions of AGN and GRBs to
the UHECR flux. For superluminal sources, such contributions can
be excluded using current neutrino flux limits to normalise the spectrum.
 In the case of subluminal sources, the
spectrum is normalised to the CR flux above the ankle,
$E_{\min}=10^{9.5}$~GeV. Using only AGN ($\Gamma=10$), the spectrum fits 
the data within experimental uncertainties. With a significant contribution from the very high relativistic shocks in GRBs
($100<\Gamma<1000$), however, the total spectrum is too flat and it is difficult to
explain the lower part of the spectrum around $\ep\sim 10^{9.5}$~GeV.
Even if UHECRs are accelerated in either external or internal shocks of GRBs, it is necessary to account
for all the energy in accelerated particles, down to the injection energy.
The total relativistic plasma output available from GRBs is only marginally sufficient to account
for the total energy required in the extra-galactic CR spectrum.

\item Recent Auger results indicate a correlation 
between CRs at the highest energies ($\ep>5.7\cdot 10^{10}$~GeV) and the
distribution of AGN \citep{auger07}. This is a first evidence that the cosmic
ray flux {\em above}
the GZK cutoff originates from AGN predominantly in the supergalactic plane. The question of
the origin of
 CRs  {\em below} the GZK cutoff is not answered by this observation, but it is likely
 that more distant AGN contribute significantly to the flux, as AGN in the
 supergalactic plane make up the flux {\em above} the GZK cutoff. Moreover the output of X-ray active AGN
within 60 Mpc provides an energy density exceeding the local estimated total extra-galactic CR energy 
density by a factor $10^{4}$. It therefore seems reasonable to believe that the 
 results from Auger can be explained by subluminal relativistic shock acceleration in AGN. It is now important
 to further develop the simulation results by including particle interactions, more detailed modelling of particle 
propagation in the inter-galactic medium and by
 investigating the source distribution of AGN in relation to observation in order
 to resolve the question of which AGN are the main sources of UHECRs.

\item Extragalactic, superluminal shocks are good candidates for the production of
  high energy neutrinos and
  photons. The energy density at low energies may be quite high
  compared to the observed flux of UHECRs, but hidden by galactic
  cosmic rays. Neutrino- and photon fluxes may, on the other hand, be
  identified and are probably the only possibility to observe
  extragalactic, superluminal shocks.
\end{enumerate}
\section*{Acknowledgements}
We are grateful to Peter Biermann, Francis Halzen and Wolfgang Rhode for extensive and fruitful 
discussions on this work. 
The project was co-funded by the European Social Fund and 
National Resources (EPEAEK II) PYTHAGORAS, Greece. 
\clearpage

\appendix
\section{The simulation \label{simulation:appendix}}
The use of a Monte Carlo technique to solve Equ.~(\ref{boltzmann:equ}) is dependent on the 
assumption that the collisions represent scattering in pitch angle and that the scattering 
is elastic in the fluid frame where there is no residual electric field. 
Since we assume that the Alfv{\'e}n waves have lower speed than the plasma
flow itself, 
the scattering is elastic in the fluid frame. A phase averaged distribution 
function is appropriate to the diffusion approximation we employ which uses many small angle scatters.

We begin the simulation by injecting $10^{5}$ particles far upstream and of a weight $w_p$ equal to 1.0. 
A splitting technique is used similar to the one used in the Monte Carlo simulations of \cite{meli_quenby03_1,meli_quenby03_2}, 
so that when an energy level is reached such that only a few accelerated 
CRs remain, each particle is replaced by a number of $N$ particles of statistical
weight $1/N$, so as to keep a roughly constant number of CRs followed.
First order Fermi (diffusive) acceleration is then simulated by following the particles' 
guidance centres and allowing for numerous pitch angle scatterings in interaction with the assumed magnetised media, while at each shock crossing the particles gain an amount of energy determined by a transformation of reference frame.
The particles are assumed to be relativistic with an initial injection energy
 of $\gamma\sim(\Gamma+100)$ when they 
are entered in the model upstream directed towards the shock.
As a justification for a test particle approach,
we note that \cite{bell78a,bell78b} and \cite{jones_ellison} have shown that
'thin' sub-shocks appear even in the non-linear regime, so at some energy above the plasma $\Gamma$
value, the accelerated particles may be dynamically unimportant while they re-cross the discontinuity.
Another way of arriving at the test-particle regime is to inject particles
well above the plasma particle energy when they are dynamically unimportant and
thus require the seed particles to have already been pre-accelerated.

\begin{figure}[h!]
\begin{center}
\epsfig{file=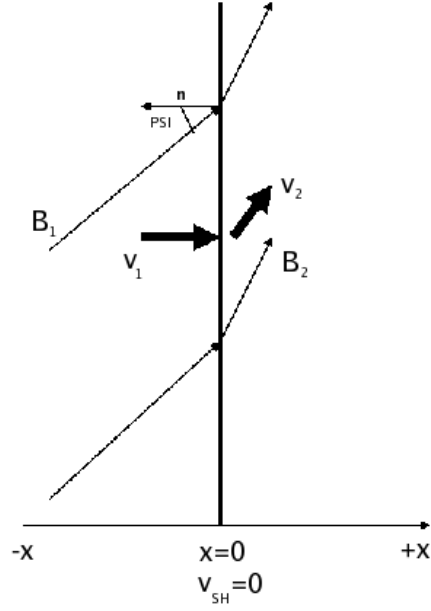,  width=7cm}
\caption{The coordinate system of a shock as seen in the so called (normal) shock frame.\label{frame_sh}}
\end{center}
\end{figure}

\begin{figure}
\begin{center}
\epsfig{file=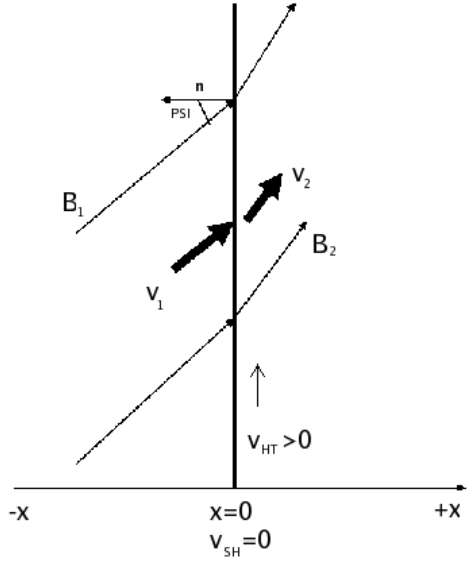,  width=8cm}
\caption{The coordinate system of a shock as seen in the so called de Hoffmann and Teller frame.\label{frame_ht}}
\end{center}
\end{figure}

The basic coordinate system employed to describe a shock is 
a Cartesian system $(x,\,y,\,z)$, where the shock plane lies on the $(y,\,z)$ plane.  
The reference frames used during the simulations are the upstream and downstream fluid
frames, the normal shock frame (NSH) and the de Hoffmann-Teller (HT) frame, see Figures~\ref{frame_sh} and~\ref{frame_ht}.


For the oblique shock cases studied here, provided the field directions encountered are 
reasonably isotropic in the shock frame, we know that
 $\tan\psi_{1}=\Gamma_{1}^{-1}\tan\psi_{NSH}\sim \Gamma_{1}^{-1}\sim \psi_{1}$
where '1' and 'NSH' refer to the upstream and normal shock frames respectively.
The concentration of field vectors close to the $x$-axis in the upstream fluid frame allows for
a reasonable probability of finding a de Hoffmann-Teller frame with a boost along the negative $y$-axis less than $c$.
Making this boost then yields an upstream HT frame inclination,
$\tan\psi_{HT,1}=\Gamma_{HT,1}\tan\psi_{1}$. While all particles are allowed to cross from downstream to upstream,
only particles with a critical HT frame pitch angle, $\theta_{c}$, given by
\begin{equation}
\theta_{c}=\arcsin (\frac{B_{HT,1}}{B_{HT,2}})^{0.5}
\label{thetac:app}
\end{equation}
are allowed to cross down to upstream and conservation of the first adiabatic invariant is used to determine the new,
downstream pitch angle. A compression ratio of 3 is used although some MHD conditions favour a value of 4.
\cite{meli_quenby03_2} do not find a considerable difference in
simulation results between these two cases.
The results of~\cite{newman1992} suggest
that it is legitimate to use Equ.~\ref{thetac:app} as a reasonable approximation. 
These authors checked the preservation of the first adiabatic invariant 
for the worst case, near perpendicular shock, employing trajectory 
integration with a realistic scattering field right up to the shock
  interface. At the critical angle for a compression ratio of three shock,
adiabatic invariant deviations were typically confined within ten percent
while large effects tended to occur only towards 90 degree pitch angle. 
Particles are assigned a random phase so that a 3-dimensional transformation of 
momentum vectors can be achieved between the fluid and HT frames.
Away from the shock, the guiding centre approximation is used so that
a test particle moving a distance, $d$, along a field line at $\psi$ to the shock normal,
in the plasma frame has a probability of collision within $d$ given by $P(d)=1-\exp(-d/\lambda)=R$,
where the random number $R$ is $0 \leq R \leq 1$. Weighting the probability by the current in the 
field direction $\mu$ yields $d=-\lambda \mu \ln R$. 
The pitch angle is measured in the local fluid frame, while the value $x_i$ gives the distance 
of the particles to the shock front, 
where the shock is assumed to be placed at $x=0$. 
Furthermore, $x_{i}$ is defined in the shock 
rest frame and the model assumes variability in only one spatial dimension.
Scattering in pitch is applied as described in the main body of this text.

In our simulations continuous Lorentz transformations are performed from and into 
the local plasma frames into or from the shock frame in order to check for particle shock crossings. 
All particles leave the system if they escape far downstream 
at the spatial boundary, $r_b$. The downstream spatial boundary required can be estimated initially
from the solution of 
the convection-diffusion equation in a non-relativistic, large-angle scattering approximation 
in the downstream plasma, which gives the chance of return to the shock as $\exp(-V_{2}r_{b}/x_i)$.
In fact, we have performed many runs with different spatial boundaries 
to investigate the effect of the size of the acceleration region on the spectrum, so as to find a 
region where the spectrum is size independent. 
Alternatively, the particles leave the system if they reach a specified maximum energy $E_{\max}$ for 
computational convenience. 

For the superluminal shock conditions, where the physical picture of the shock drift acceleration applies,
it is necessary to abandon the guiding centre approximation when the trajectories begin to intersect 
the shock surface. Here, we consider a helical trajectory motion of each test-particle of momentum $p$,  
in the fluid frame, upstream or downstream, where the velocity coordinates ($v_x, v_y, v_z$) 
of the particle are calculated in 3-dimensional space as follows

\begin{equation}
\upsilon_{x_i}=\upsilon_i \cos\theta_i \cos\psi_i-\upsilon_i \sin\theta_i \cos\phi_i \sin\psi_i\,,
\end{equation}
 
\begin{equation}
\upsilon_{y_i}=\upsilon_i \cos\theta_i \sin\psi_i+\upsilon_i \sin\theta_i \cos\phi_i \cos\psi_i
\end{equation}
and
\begin{equation}
\upsilon_{z_i}=-\upsilon_i \sin\theta_i \sin\phi_i\,.
\end{equation}
Here, $\theta_i$ is the pitch angle, $\phi_i\in(0,2\pi)$ and $\psi_i$ is the angle between the magnetic field
and the shock normal in the respective fluid frames ($i$=1,2 for upstream and downstream respectively).
 
We follow the trajectory in time, using $\phi_i=\phi_{\circ}+\omega t$, where
 $t$ is the time from detecting shock presence at $x_{NSH}$, $y_{NSH}$, $z_{NSH}$ by using
\begin{equation}
dx=x_{NSH}+\upsilon_{x_i} \delta t\,,
\end{equation}
\begin{equation}
dy=y_{NSH}+\upsilon_{y_i} \delta t
\end{equation}
and
\begin{equation}
dz=z_{NSH}+\upsilon_{z_i} \delta t\
\end{equation}
assuming that $\delta t=r_{g}/Hc$ where $H \ge 100$ and $r_{g}$ is the Larmor radius.
The particle's gyrofrequency $\omega$ is given by the relation,
$\omega_i=e |\vec{B}_i|/\gamma_i$, $\vec{B}_i$ is the magnetic field, $\gamma_i$ is the particle's
boost factor and $e$ is its charge in gaussian units.

We follow the helical trajectory of each particle in time $t$, in the new frame
where $t$ is the time from detecting the shock intersection at $(x,\,y,\,z)$ until the trajectory
has performed one gyro period without re-intersecting the shock surface. 
Nevertheless, because of the peculiar properties of the helix we need to establish 
where a particle, starting off in the upstream frame, with a particular $\theta$ and $\phi$ first
encounters the shock. To establish when the shock encounter happens, we choose to go back a whole period,
$T_i=2\pi/\omega_i$ by reversing signs of the helix velocity coordinates and by keep checking 
throughout the simulation to determine if the particle trajectory encounters the shock front, placed at 
$x=0$ in the shock rest frame. 
If the particle encounters the shock then the suitable Lorentz transformation to 
the relevant fluid rest frame is made and we continue following the particle helical trajectory
until shock intersections cease. At this juncture, the guiding centre is followed in 
the same way as in the diffusive acceleration picture of the subluminal shocks.
During the helical phase of the numerical integration, the prescription for pitch angle scatter
is applied as in the general plasma frame motions, in order to more realistically simulate a mean plus 
chaotic field situation where turbulence is clearly present close to a shock. 
\section{The spectra\label{spectra:appendix}}
\begin{figure}[h!]
\begin{center}
\epsfig{file=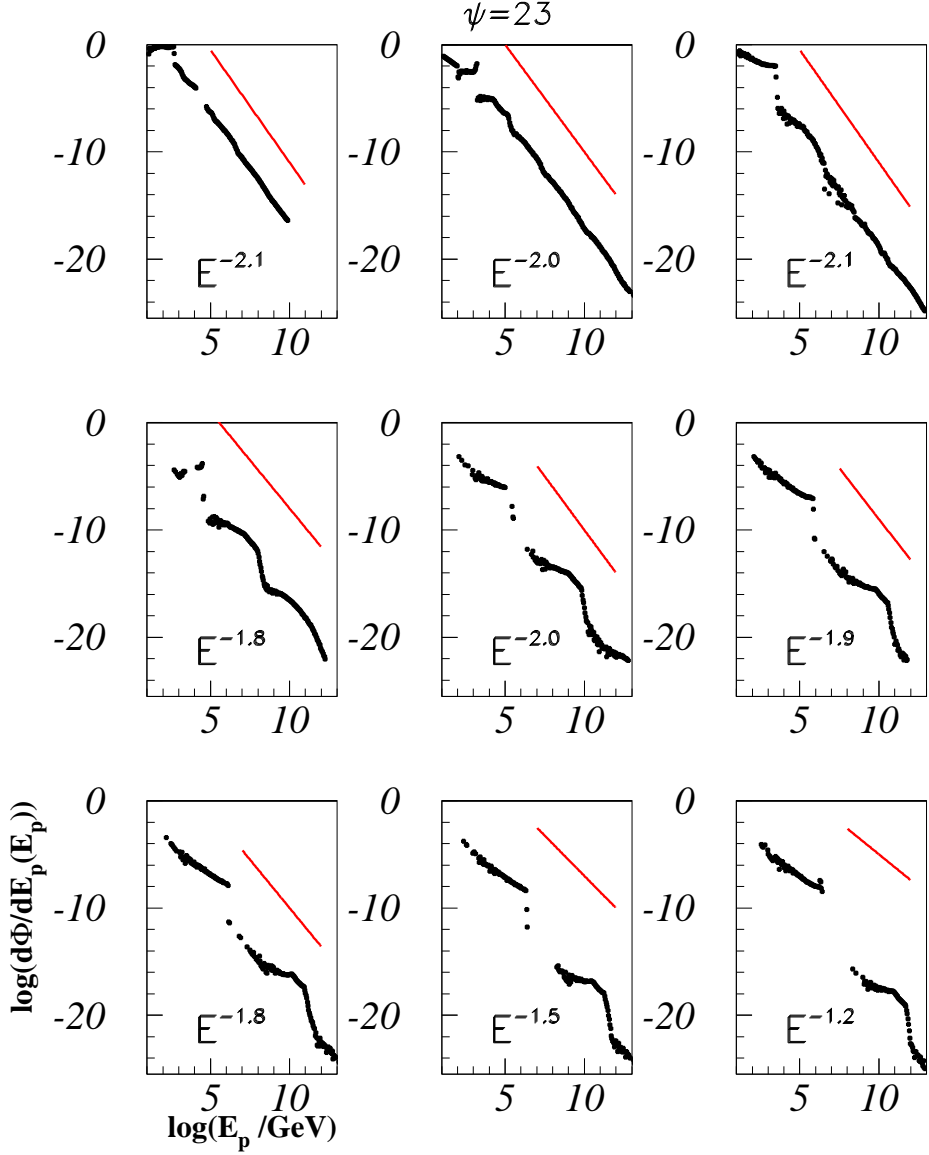, width=\linewidth}
\caption[Subluminal spectra for $\psi=23^{\circ}$]{Subluminal spectra for $\psi=23^{\circ}$ and different
  $\Gamma$: $\Gamma=10,20,30$ is displayed in the first row, in the middle,
  $\Gamma=100,300,500$ is shown and $\Gamma=700,900,1000$ is the bottom
  row. The black dots in each graph represent the simulation result. The straight
  line shows the single power-law for comparison. The spectral behaviour is
  indicated in the lower left corner of each graph.}
\label{subl_spectra_23}
\end{center}
\end{figure}

\begin{figure}[h!]
\begin{center}
\epsfig{file=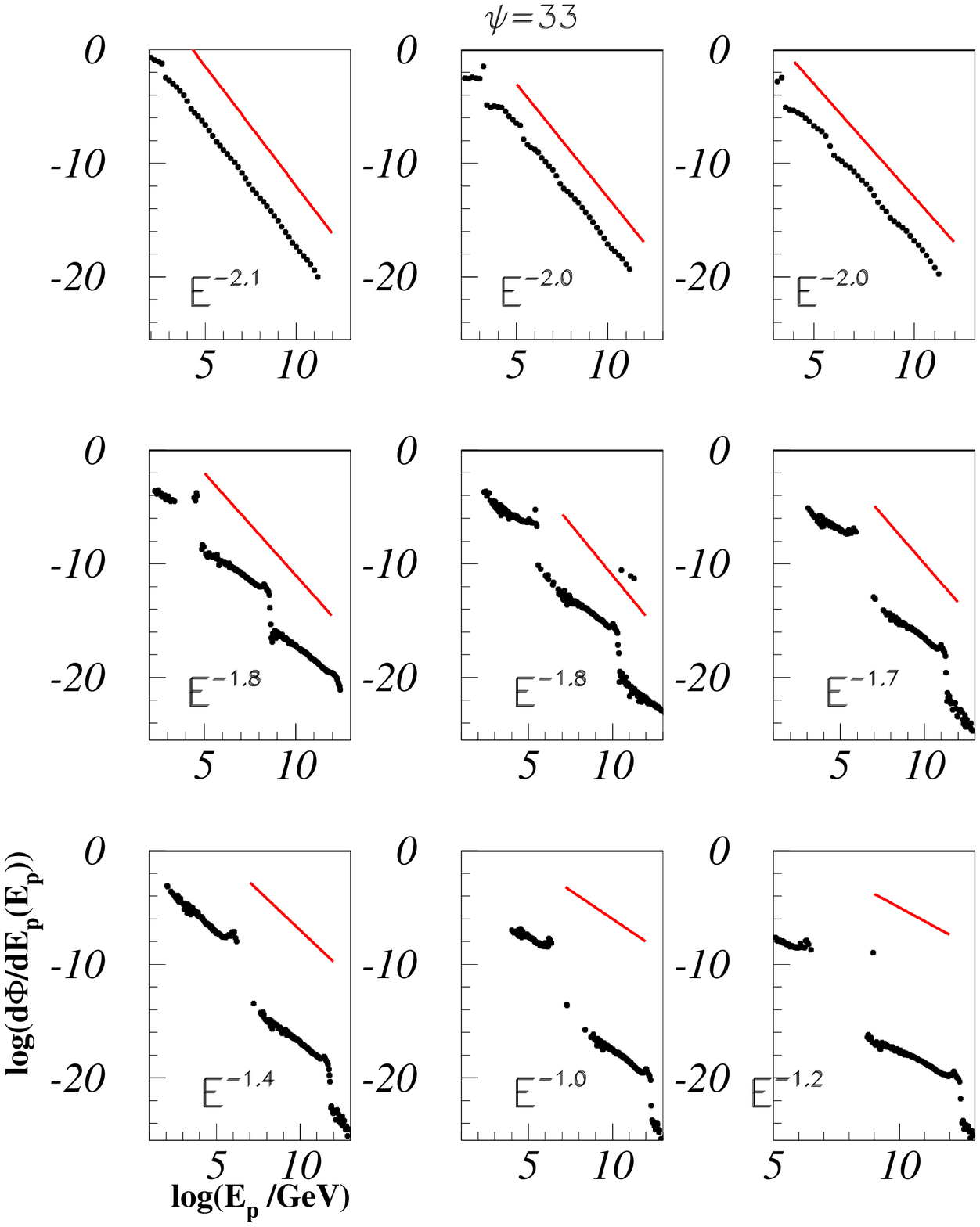, width=\linewidth}
\caption[Subluminal spectra for $\psi=33^{\circ}$]{Subluminal spectra for $\psi=33^{\circ}$ and different
  $\Gamma$, as in Fig.~\ref{subl_spectra_23}.}
\label{subl_spectra_33}
\end{center}
\end{figure}

\begin{figure}[h!]
\begin{center}
\epsfig{file=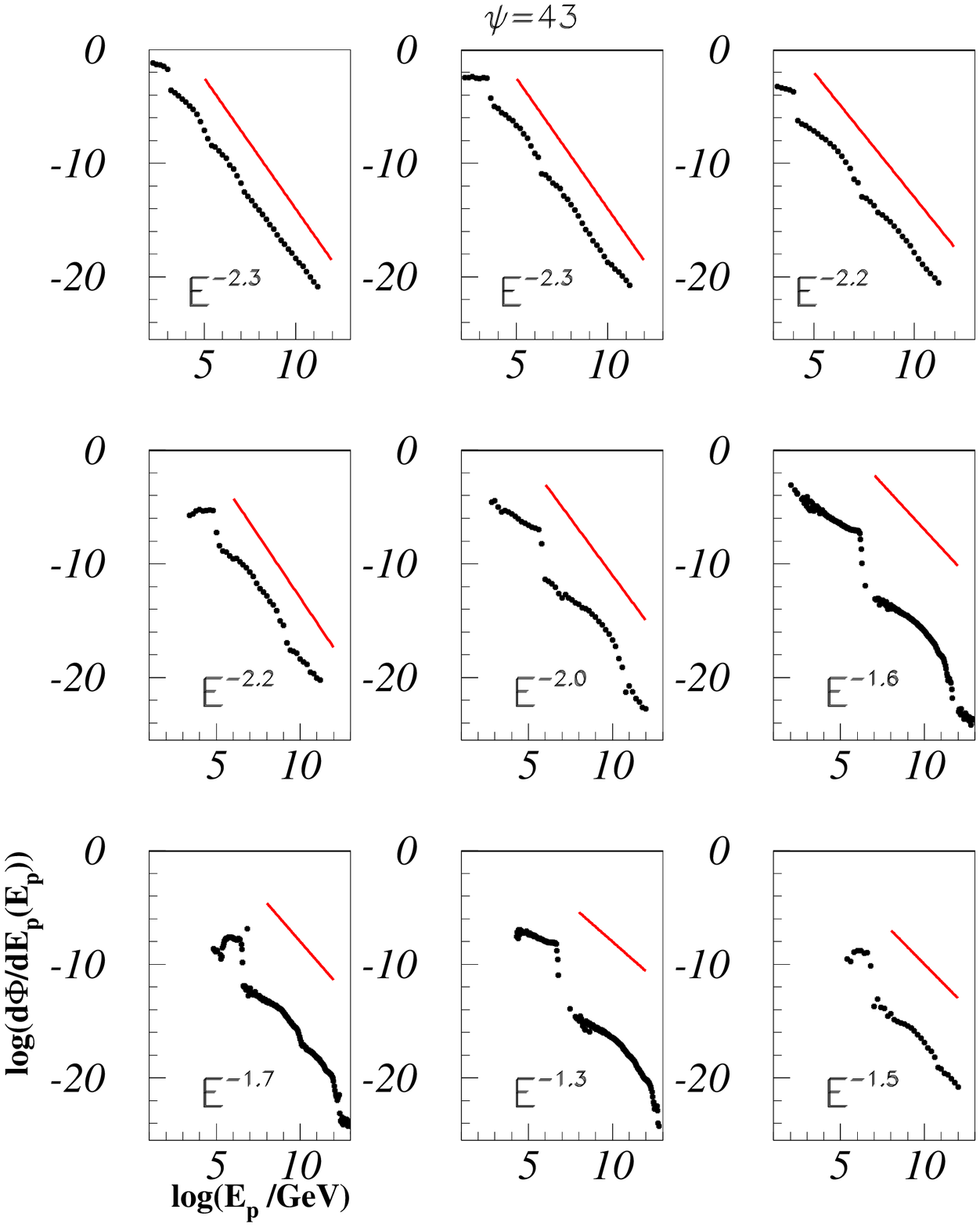, width=\linewidth}
\caption[Subluminal spectra for $\psi=43^{\circ}$]{Subluminal spectra for $\psi=43^{\circ}$ and different
  $\Gamma$, as in Fig.~\ref{subl_spectra_23}.}
\label{subl_spectra_43}
\end{center}
\end{figure}

The particle spectra resulting from the simulation are presented in
Figures~\ref{subl_spectra_23}, \ref{subl_spectra_33} and \ref{subl_spectra_43}.
In each of the figures, the simulated particle spectra are shown (black dots) for three different shock angles each,
$\psi=23^{\circ} , 33^{\circ}, 43^{\circ}$, for nine different boost factors, $\Gamma=10,20,30$ in the first row, 
starting from the left, $\Gamma=100,300,500$ in the middle row and $\Gamma=700,900,1000$ in the lower
row. 
Each graph shows the logarithm of the proton spectrum $d\Phi/d\ep$ in
arbitrary units versus the proton's energy in units of GeV. Note that the spectrum can 
be expressed more generally  in terms of the particle's boost factor $\gamma=\ep/(m_p\,c^2)$. 
Therefore, the results are also valid for nuclei with higher mass (e.g. Fe).
In the present investigation protons are considered. 

Figures~\ref{subl_spectra_23},~\ref{subl_spectra_33} and~\ref{subl_spectra_43} show that the spectral shape 
starts to deviate from a power-law as $\Gamma$ increases with the onset of plateau formation. This might be expected, since  the
particles are swept away rapidly downstream with a low possibility to return upstream
 for high $\Gamma$ factors (around 20\% of the particles
return to the shock after one shock cycle). A small chance to return to the shock (except 
for a relatively small subset of downstream pitch angle particle 'histories') produces the spectral 
irregularities and the anisotropy  seen for these returning particles.
We note that in the cases of $\Gamma=10-30$
relatively smooth spectra are produced, but spectra become more structured (plateau-like) at 
the more extreme values $\Gamma\rightarrow 1000$.  In the latter cases the effects of individual 
acceleration cycles are clearly evident. 
The mechanism of plateau development as an acceleration cycle effect is
implicit in figure 6 of \cite{Protheroe01}  and figure 2 of~\cite{Steckeretal07} and seems to be independent 
of the shock inclination angle for a particular scattering model. The structured spectra can also be seen in the lower $\Gamma$ 
large angle scatter model simulations of \cite{QuenbyLieu89} and \cite{Ellisonetal90} while \cite{Protheroe01} shows a 
similar contrast in behaviour, between large and small angle scattering models up to $\Gamma=20$.

The initial bump in most of the spectra  is due to the
monochromatic injection of the particles. We approximate
the spectra with a power-law (straight lines seen in the figures) at higher energies and thus,
artificial injection features are excluded from the calculations. The
maximum energy is chosen to be $10^{21}$~eV as discussed in Section~\ref{emax}.
The plateau-like parts in the spectra at higher energies are physical features, especially at high boost
factors, as particles continue to be accelerated in a second cycle. The lower energy part
of the spectrum is dominated by particles undergoing one acceleration shock cycle,
while the second bump in the spectrum represents particles having completed two
cycles. A shock cycle is a shock crossing from upstream to downstream to upstream. Since some upstream particles
can suffer reflection before downstream transmission, this effect increases the statistical energy gain
 of the particles in their overall encounter with the shock surface.
The aim of fitting a single power-law to the computed spectral points is primarily to examine the
variation of primary spectra with the  $\Gamma$ factor of the shock. The spectral points themselves are
difficult to compare because the structure becomes more complex with the increasing
boost factor. 
As discussed in the main text, the relevant way to compare the
particle spectra to the data is to fit the energy range observed in cosmic
rays. The straight lines in Figures~\ref{subl_spectra_23},
\ref{subl_spectra_33} and \ref{subl_spectra_43} together with the
values written in the lower left corner of each graph indicate the single power-law
for comparison chosen individually for each case.

Table~\ref{gamma_index:tab}
shows the spectral index dependence of the boost factor for the three shock inclination angles
$\psi=23^\circ,33^{\circ},43^{\circ}$. For each of the three angles, the spectra become harder
with the increasing $\Gamma$ factor of the shock. Due to the structure in the spectra, these
values represent simple first order approximations which are useful for 
comparison of the simulation results with the data. CR data include
large statistical and systematic errors, which would make it difficult to
distinguish features attributable to the acceleration mechanism other than single or broken power-laws.


The spectral flattening with higher $\Gamma$, implies that 
GRB particle spectra, arising from relativistic shocks
with very high boost factors between $100<\Gamma<1000$, have spectral indices ranging between 
$\alpha_p\sim 2.1 - 1.5$. These spectra tend to be somewhat steeper than some particular entries in Table~\ref{gamma_index:tab},
but the tendency of a flattening with the boost factor is always present. The relativistic flattening effect is 
consistent with  studies of \cite{Baring99, baring04b, Steckeretal07}. 
On the other hand, for shocks with $\Gamma\sim 10-30$ occurring in AGN, 
spectral indices have values between $ 2.0 < \alpha_p < 2.3$. 

In addition, as one sees from Table~\ref{gamma_index:tab}, the single power-law comparison gives comparable values for
different angles $\psi$. Test simulation runs we performed, using a series of values of $\psi$, follow
the same trend as long as the scattering model is fixed.
The present results deviate from earlier investigations by \cite{kirk00, ostrowski_bednarz02}, which 
show a saturation of the hardening of the spectra with the boost factor, 
$\alpha_p\rightarrow 2.33$ for $\Gamma\rightarrow \infty$. Nevertheless, the latter studies concern 
the special cases of extremely small values for pitch angle scattering, for particle acceleration in relativistic shocks.
\begin{table}
\centering{
\begin{tabular}{|ll|l|l|}
\hline
$\Gamma$&$\alpha_p(\psi=23\deg)$&$\alpha_p(\psi=33\deg)$&$\alpha_p(\psi=43\deg)$\\\hline\hline
10&2.1&2.1&2.3\\\hline
20&2.0&2.0&2.3\\\hline
30&2.1&2.0&2.2\\\hline
100&1.8&1.8&2.2\\\hline
300&2.0&1.8&2.0\\\hline
500&1.9&1.7&1.6\\\hline
700&1.8&1.4&1.7\\\hline
900&1.5&1.0&1.3\\\hline
1000&1.2&1.2&1.5\\\hline
\end{tabular}
\caption{Spectral indices for a single power-law comparison for subluminal
 shocks of different boost factors and three inclination angles.\label{gamma_index:tab}}
}
\end{table}

\clearpage
\bibliography{relativistic_super_sub_Mar3JKB.bib}
\bibliographystyle{elsart-harv}


\end{document}